\documentclass[english,aps,prb,superscriptaddress,twocolumn,amsmath,amssymb]{revtex4-1}
\usepackage{graphicx}
\usepackage{multirow}
\usepackage{epstopdf}
\usepackage{gensymb}
\usepackage{color}
\usepackage{amsmath}
\usepackage{array}
\usepackage{amsmath,amsfonts,amsthm,bm} 
\usepackage{fixltx2e} 

\begin{document}

\title{Precise control of $J_\mathrm{eff}$ = $\frac{1}{2}$ magnetic properties in Sr$_2$IrO$_4$ epitaxial thin films by variation of strain and thin film thickness}

\author{Stephan Gepr\"ags}
	\email[]{stephan.gepraegs@wmi.badw.de}
	\affiliation{Walther-Mei{\ss}ner-Institut, Bayerische Akademie der Wissenschaften, Garching, Germany}
\author{Bj\"orn Erik Skovdal}
	\affiliation{Division of Synchrotron Radiation, Lund University, Lund, Sweden}
	\affiliation{Angstrom Centre, Uppsala University, Sweden}
\author{Monika Scheufele}
	\affiliation{Walther-Mei{\ss}ner-Institut, Bayerische Akademie der Wissenschaften, Garching, Germany}
	\affiliation{Physik-Department, Technische Universit\"{a}t M\"{u}nchen, Garching, Germany}
\author{Matthias Opel}
	\affiliation{Walther-Mei{\ss}ner-Institut, Bayerische Akademie der Wissenschaften, Garching, Germany}
\author{Didier Wermeille}
	\affiliation{XMaS CRG Beamline, European Synchrotron Radiation Facility, Grenoble, France}
\author{Paul Thompson}
	\affiliation{XMaS CRG Beamline, European Synchrotron Radiation Facility, Grenoble, France}
\author{Alessandro Bombardi}
	\affiliation{Diamond Light Source Ltd., Harwell Science \& Innovation Campus, Didcot, United Kingdom}
\author{Virginie Simonet}
	\affiliation{Univ. Grenoble Alpes, CNRS, Grenoble INP, Institut N$\acute{e}$el, 38000 Grenoble, France}
\author{St$\acute{e}$phane Grenier}
	\affiliation{Univ. Grenoble Alpes, CNRS, Grenoble INP, Institut N$\acute{e}$el, 38000 Grenoble, France}
\author{Pascal Lejay}
	\affiliation{Univ. Grenoble Alpes, CNRS, Grenoble INP, Institut N$\acute{e}$el, 38000 Grenoble, France}
\author{Gilbert Andre Chahine}
	\affiliation{Univ. Grenoble Alpes, CNRS, Grenoble INP, SIMaP, 38000 Grenoble, France}
\author{Diana Quintero Castro}
	\affiliation{Department of Mathematics and Physics, University of Stavanger, Stavanger, Norway}
\author{Rudolf~Gross}
  \affiliation{Walther-Mei{\ss}ner-Institut, Bayerische Akademie der Wissenschaften, Garching, Germany}
  \affiliation{Physik-Department, Technische Universit\"{a}t M\"{u}nchen, Garching, Germany}
  \affiliation{Munich Center for Quantum Science and Technology (MCQST), Munich, Germany}
\author{Dan Mannix}
\email[]{dan.mannix@esss.se}
	\affiliation{Univ. Grenoble Alpes, CNRS, Grenoble INP, Institut N$\acute{e}$el, 38000 Grenoble, France}
	\affiliation{European Spallation Source, SE-221 00 Lund, Sweden}
	\affiliation{Aarhus University, Langelandsgade 140, DK-8000 Aarhus, Denmark}

\begin{abstract}
We report on a comprehensive investigation of the effects of strain and film thickness on the structural and magnetic properties of epitaxial thin films of the prototypal $J_\mathrm{eff} = 1/2$ compound Sr$_2$IrO$_4$ by advanced X-ray scattering. We find that the Sr$_2$IrO$_4$ thin films can be grown fully strained up to a thickness of 108\,nm. By using X-ray resonant scattering, we show that the out-of-plane magnetic correlation length is strongly dependent on the thin film thickness, but independent of the strain state of the thin films. This can be used as a finely tuned dial to adjust the out-of-plane magnetic correlation length and transform the magnetic anisotropy from two-dimensional ($2D$) to three-dimensional ($3D$) behavior by incrementing film thickness. These results provide a clearer picture for the systematic control of the magnetic degrees of freedom in epitaxial thin films of Sr$_2$IrO$_4$ and bring to light the potential for a rich playground to explore the physics of 5$d$ transition metal compounds. 
\end{abstract}

\maketitle

\section{Introduction}
\label{sec:intro}

The complex 3$d$ transition metal oxides have emerged as one of the most fascinating and technologically relevant materials systems, with a plethora of physical properties,\cite{Rao:1989} including orbital physics,\cite{Tokura:2000} multiferroicity,\cite{Wang:2009} metal-insulator transitions,\cite{Imada:1998} colossal magnetoresistance,\cite{Urushibara:1995} as well as high temperature superconductivity.\cite{Bednorz:1986} The 5$d$ transition metal compounds offer the potential to further enrich the physics of this class of materials by the possibility of generating new physical properties within the strong spin-orbit limit.\cite{Pesin:2010} Unfortunately, strongly correlated 5$d$ transition metal compounds are rather scarce compared to their 3$d$ counterparts. This greatly reduces the opportunities to explore the science potential of these compounds. However, the 5$d$-iridates of the Ruddlesden-Popper series Sr$_{n+1}$Ir$_n$O$_{3n+1}$ have attracted considerable scientific attention due to the possibility of generating  spin-orbit driven Mott insulating states.\cite{Jackeli:2009,Wan:2011,Krempa:2014} In the first member of the series, the layered Sr$_2$IrO$_4$ (SIO) ($n=1$) compound, the large spin-orbit coupling combined with a large crystal field splitting results in a Mott insulating ground state, in which the local electronic state with nominally $(t_{2g})^5$ electron configuration of the Ir$^{4+}$-ions is represented by an effective total angular momentum $J_\mathrm{eff}$ = $1/2$.\cite{Kim:2008,Moon:2008,Kim:2009,Lu:2020} Below 240\,K,\cite{Crawford:1994} this pseudo-$1/2$-spin orders antiferromagnetically in a layered magnetic structure akin to that of the seminal cuprate high-$T_c$ superconducting parent La$_2$CuO$_4$ (LCO) compound. The structural and magnetic similarities to LCO, including the emergence of a pseudogap,\cite{Kim:2014,Yan:2015,Battisti:2016} a $d$-wave gap at low temperature,\cite{Kim:2016} as well as a hidden non-dipolar magnetic order,\cite{Zhao:2015,Jeong:2017} has led to considerable research on possible unconventional superconductivity in SIO.\cite{Qi:2011,Watanabe:2013,Meng:2014} 

Here, we report on advanced X-ray resonant scattering (XRS) studies on epitaxial films of SIO in a thickness range of 27\,nm to 108\,nm. We found a strong dependence of the magnetic out-of-plane correlation length on the thickness of the SIO thin films independent of the strain state. Furthermore, only a weak intensity of reflections related to the canting of the Ir-moments could be found, which might be related to a disordered oxygen-lattice structure in the SIO thin films. Our results reveal a vast potential for fine tuning the magnetic properties of SIO thin films, which could provide new routes into the systematic investigation of 5$d$-Mott insulators with strong spin-orbit-coupling.

The paper is organized as follows: In Sec.~\ref{sec:Physical-Properties}, we briefly review the physical properties of SIO reported for bulk and thin film materials. In the following section (Sec.~\ref{sec:Preparation-Characterization}), we describe the fabrication and initial structural and magnetic characterization using X-ray diffraction and SQUID magnetometry. Section~\ref{sec:A-sublattice} presents our XRS studies of the magnetic $A$-sublattice, while Sec.~\ref{sec:B-sublattice} reports on studies aimed at investigating the magnetic $B$-sublattice of SIO, which arises from the canting of the Ir$^{4+}$-moments in SIO. Finally, Sec.~\ref{sec:conclusion} presents the conclusions of our investigations.   

\section{Physical properties of S\lowercase{r}\textsubscript{2}I\lowercase{r}O\textsubscript{4}}
\label{sec:Physical-Properties}

\begin{figure}
\centering
\includegraphics[width=\columnwidth]{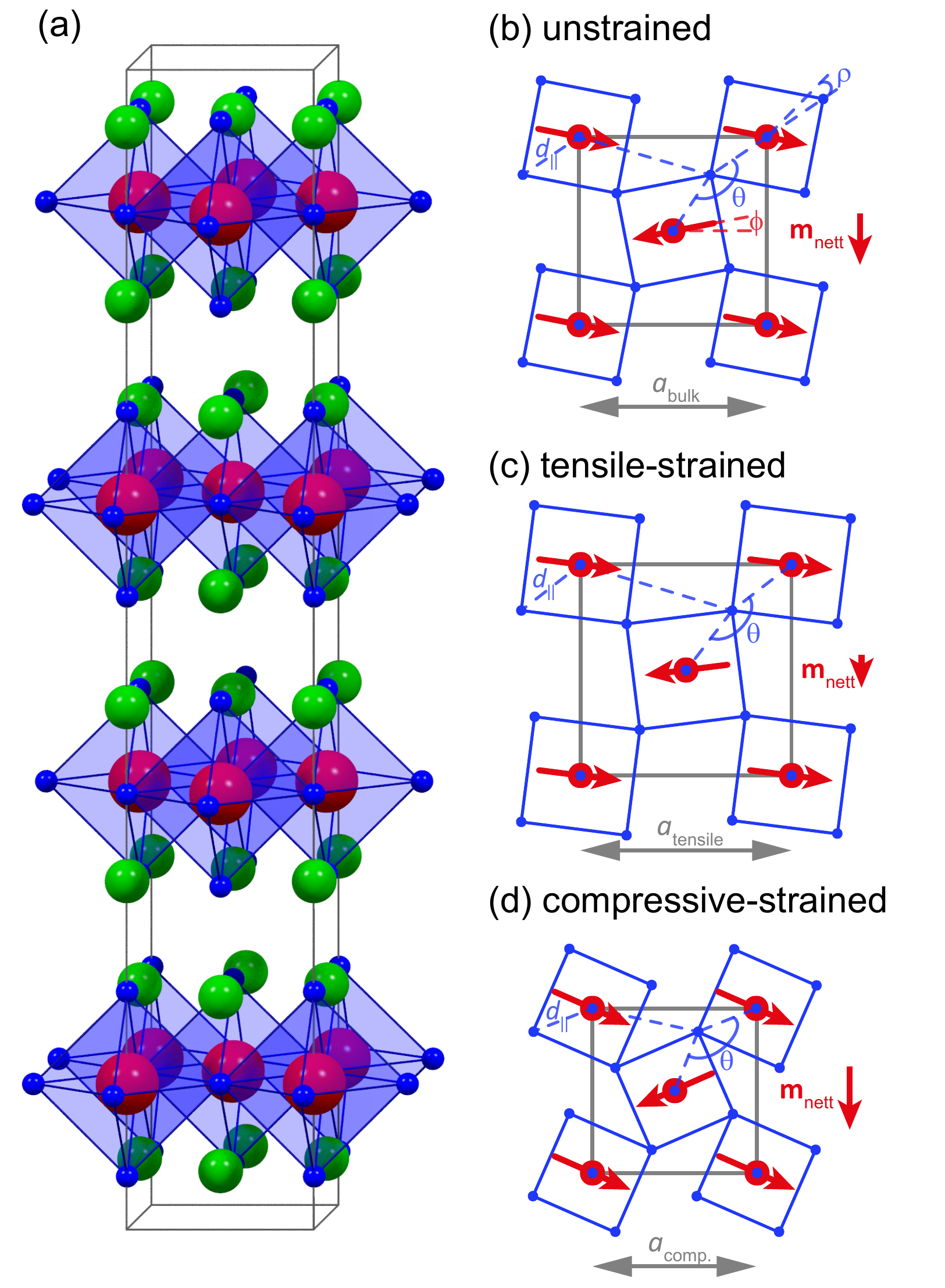}
\caption{(a) Schematic of the Sr$_{2}$IrO$_{4}$ unit cell. Blue spheres represent O$^{2-}$, red spheres Ir$^{4+}$, and green spheres Sr$^{2+}$ ions. The oxygen octahedra are highlighted. (b)-(d) Schematic representation of the expected structural changes of the IrO$_2$ planes under (c) tensile and (d) compressive strain compared to the (b) unstrained, bulk case. From the Ir-O bond length $d_\parallel$ and the Ir-O-Ir bond angle $\theta$, the in-plane lattice constant $a$ can be calculated from $2d_\parallel\sqrt{1-\cos(\theta)}$.\cite{Nichols:2013} The red arrows display the magnetic moments of the Ir$^{4+}$-ions. The finite tilting of the moments with tilting angle $\theta$ results in a net magnetic moment $\mathbf{m}_\mathrm{net}$.}  
\label{fig:fig1}
\end{figure}

In bulk form, SIO crystallizes in the tetragonal structure. The space group $I4_1/adc$ with lattice parameters $a=b=5.48$\,\AA\,and $c=25.8$\,\AA\,has been resolved by neutron powder diffraction.\cite{Crawford:1994} However, recently, a small displacement of the planar oxygen atoms resulting in additional, weak $(1\,0\,2n+1)$-type reflections in neutron diffraction experiments has been found,\cite{Dhital:2013} which is explained by a symmetry reduction from $I4_1/adc$ to $I4_1/a$.\cite{Ye:2015} This is further confirmed by nonlinear optical harmonic generation experiments.\cite{Torchinsky:2015} In either case, the SIO unit cell is composed of four IrO$_2$ layers separated by Sr$^{2+}$-ions along the $c$-axis (see Fig.~\ref{fig:fig1}(a)). Within the IrO$_2$ layers in the $ab$-plane, the $\mathrm{Ir}^{4+}$ ions are centered in elongated oxygen octahedra, which are alternately rotated by $\rho=11.8^{\circ}$ about the $c$-axis with respect to the ideal $I4/mmm$ tetragonal space group (cf.~Fig.\ref{fig:fig1}(a)).\cite{Crawford:1994,Boseggia:2013,Ye:2013} Below the N\'eel temperature of around $T_N=240$\,K,\cite{Crawford:1994,Shimura:1995,Cao:1998} the Ir$^{4+}$-spins and coupled orbital moments\cite{Fujiyama:2014} follow the rotation of the oxygen octahedra and order in a canted antiferromagnetic (AFM) structure with a next nearest-neighbor exchange constant $J=60$\,meV\cite{Kim:2012} and a canting angle of the Ir$^{4+}$-moments of $\phi = 12.2^{\circ}$-$13.0^{\circ}$ (cf.~Fig.\ref{fig:fig1}(a)).\cite{Boseggia:2013,Ye:2013} The locking of the Ir$^{4+}$-moments with respect to the rotation of the oxygen octahedra was first predicted by Jackeli and Khaliullin (JK).\cite{Jackeli:2009} Within this JK-model, the ratio $\phi/\rho$ depends on the strength of the spin-orbit coupling $\lambda$ and the tetragonal crystal field splitting $\Delta$ and can be calculated to $\phi/\rho \approx 0.7$ by using $\lambda \approx 400$\,meV and $\Delta \approx 140$\,meV.\cite{Torchinsky:2015}  The discrepancy with the almost perfect magnetoelastic locking ($\phi/\rho \approx 1$) derived from experiment might be caused by the presence of unequal tetragonal distortions within the $I4_1/a$ space group,\cite{Ye:2013} which was taken into account in a modified JK-model.\cite{Torchinsky:2015} However, the canting of the Ir$^{4+}$-moments results in a weak net magnetic moment $\mathbf{m}_\mathrm{net}$ of around $0.048\,\mu_\mathrm{B}$ along the $b$-axis (cf.~Fig.\ref{fig:fig1}(b)).\cite{Ye:2013} Due to the layered-structure of SIO, different stacking sequences along the $c$-axis of the net moment $\mathbf{m}_\mathrm{net}$ have been discussed recently,\cite{Mitchell:2015,Matteo:2016,Sumita:2017,Porras:2019} which is mainly the result of the weak inter-plane magnetic exchange in SIO of $\approx 1\,\mu$eV.\cite{Kim:2012,Fujiyama:2012,Vale:2015} In the following, we will refer to the commonly accepted stacking sequence "up-down-down-up" labeled as $uddu$.\cite{Kim:2009,Boseggia:2013,Dhital:2013,Porras:2019} Therefore, the magnetic structure of SIO can be decomposed into a basal-plane AFM sublattice $A$ with magnetic moments $m_a=0.202\,\mu_\mathrm{B}$ along the $a$-axis and a magnetic sublattice $B$ with net magnetic moments $m_\mathrm{net}=0.048\,\mu_\mathrm{B}$. In total, a magnetic moment of $m_\mathrm{total}=0.208$-$0.36\,\mu_\mathrm{B}/\mathrm{Ir}$ was derived from neutron scattering experiments.\cite{Lovesey:2012,Dhital:2013,Ye:2013} Above a critical magnetic field of around 200\,mT, a metamagnetic transition occurs, resulting in a ferromagnetic alignment of the net magnetic moments $\mathbf{m}_\mathrm{net}$ with a stacking sequence $uuuu$ along the $c$-axis.\cite{Kim:2009,Porras:2019}

Therefore, the magnetic structure of SIO is very similar to that of the superconducting parent compound LCO, but with some subtle important differences. In LCO, the spin structure is also canted, but the canting is out-of-plane (along the $c$-axis) rather than within the $ab$-plane as in SIO with a much smaller tilting angle of $\rho = 0.17^{\circ}$.\cite{Kastner:1998} This smaller spin canting angle results in a smaller net magnetic moment of $\sim 0.002\,\mu_\mathrm{B}/\mathrm{Cu}$.\cite{Kastner:1998} Therefore, SIO and LCO could represent model systems to investigate the physical effects which promote or suppress superconductivity in oxides materials, especially if these materials can be manipulated to mimic each other more closely using external perturbations, such as doping\cite{Yan:2015}, pressure\cite{Haskel:2012} or elastic strain.\cite{Nichols:2013,Serrao:2013,Lupascu:2014,Miao:2014,Liu:2015,Kim:2016a,Kim:2017,Llorente:2018,Bhandari:2018,Seo:2019} In particular, the latter could be a promising method, due to the strong magnetoelastic locking of the canting of Ir-moments on the rotation of the oxygen octahedra. In heteroepitaxy using substrates with large lattice mismatch with respect to the thin film material, it is energetically favorable for the film to accommodate the substrate-imposed epitaxial strain by the formation of defects, such as misfit dislocations or crystallographic domain structures.\cite{Carbone:2004,Dumont:2009} However, below a critical thickness $d_c$, a pseudomorphic growth on heterogeneous substrates is possible, where the misfit strain will lead to modifications of the in-plane ($d_\parallel$) and out-of-plane ($d_\bot$) bond lengths as well as the Ir-O-Ir bond angle $\theta$, i.e. the tetragonality $d_\bot/d_\parallel$ as well as the rotation angle $\rho$ of the IrO$_6$ octahedra. Tensile (compressive) strain is expected to increase (decrease) both the rotation angle $\rho$ and the Ir-O bond length $d_\parallel$ (cf.~Fig.~\ref{fig:fig1}(b)-(c)).\cite{Kim:2016a,Bhandari:2018} As discussed by Kim \textit{et al.}, this mainly leads to an enhancement of the Coulomb correlation $U$, the spin-orbit interaction $\lambda$,\cite{Kim:2016a} and to an increase of the magnetic moment upon strain.\cite{Bhandari:2018} For larger changes of $d_\bot/d_\parallel$ and $\rho$, ab initio calculations reveal changes of the canted-AFM ground state towards a collinear-AFM ordering and even a reorientation of the Ir-moments from within the $ab$-plane to the out-of-plane direction along the $c$-axis.\cite{Liu:2015} While theoretically the bond lengths as well as the bond angles change upon strain, X-ray resonant magnetic scattering studies on tensile and compressive strained 50\,nm thick epitaxial SIO films suggested that the modification of the bond length $d$ provides the major crystallographic change to accommodate epitaxial strain in SIO, leading to an increase (decrease) of the N\'eel temperature of about 30\,K of tensile (compressive) strained SIO thin films. This was ascribed by a change in inter-layer magnetic coupling by epitaxial strain.\cite{Lupascu:2014} However, this is in contradiction to recent studies using Raman spectroscopy on tensile and compressive strained SIO thin films.\cite{Seo:2019} Here, a decrease of the N\'eel temperature of about 10\,K was found in tensile strained SIO thin films. Furthermore, a shift of the two-magnon peaks to higher energies was observed as compressive strain was applied to the SIO thin film. This corresponds to an increase of the coupling strength $J$ and hopping integral $t$ in compressive strained SIO thin films and points to the multi-orbital characteristics of the $J_\mathrm{eff}$ = $1/2$ wave function of SIO.   

To further elucidate the effect of strain and size on the magnetic properties of SIO, we present here the results of an advanced XRS study on SIO thin films with thickness between 27\,nm and 108\,nm. In the following, we first discuss the fabrication and the structural properties of tensile and compressive strained epitaxial SIO thin films.

\section{Sample fabrication and characterization}
\label{sec:Preparation-Characterization}

\subsection{Sample fabrication}

Tensile and compressive strained epitaxial SIO thin films with thickness between 27-108\,nm were fabricated on (001)-oriented SrTiO$_{3}$ (STO) and (110)-oriented NdGaO$_{3}$ (NGO) substrates by pulsed laser deposition (PLD) monitored by \textit{in-situ} reflection high-energy electron diffraction (RHEED).\cite{Opel:2013} The deposition was carried out in an oxygen atmosphere with a pressure of $25\,\mu$bar, a repetition rate of the laser of 2\,Hz, and a laser fluence at the target surface of 2\,J/cm$^2$, using a stoichiometric, polycrystalline SIO target. For the fabrication of a 96\,nm thick SIO thin film on STO, a SIO target with an Ir-excess of around 30\% was used to compensate possible Ir-losses during the deposition.

\begin{figure}
\centering
\includegraphics[width=\columnwidth]{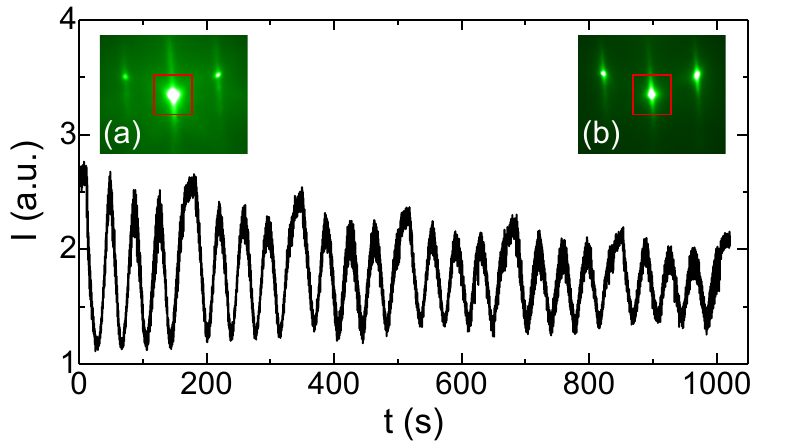}
\caption{Intensity evolution of the RHEED (0\,0) reflection monitored during the deposition of the last 6 unit cells of a 108\,nm thick SIO thin film on a STO substrate. The intensity was integrated within the red rectangle marked in the RHEED pattern shown in the insets: (a) RHEED pattern of the STO substrate before the deposition, (b) RHEED pattern after the deposition of SIO.
}  
\label{fig:fig2}
\end{figure}

The growth process of SIO is exemplary illustrated in Fig.~\ref{fig:fig2} on the basis of the RHEED intensity evolution of the (0\,0) reflection, which was recorded during the growth of the last 6 unit cells of a 108\,nm thick SIO thin film on a STO substrate. Before the deposition, the RHEED pattern of the STO substrate reveals three sharp spots of the first Laue circle, which can be indexed with ($\overline{1}$\,0), (0\,0), and (1\,0) (see inset (a) in Fig.~\ref{fig:fig2}). During the deposition, the RHEED pattern changes only marginally and RHEED oscillations are observed until the end of the deposition (see inset (b) in Fig.~\ref{fig:fig2}). This demonstrates a two-dimensional layer-by-layer growth with a smooth surface of the SIO thin film. The growth of one unit cell of SIO is manifested by four RHEED oscillations, indicating a block-by-block growth mode of the four charge neutral blocks of the unit cell.\cite{Gross:2000} We therefore divided the whole growth process into the deposition of one unit cell, followed by a growth interruption of 20\,s, allowing for the relaxation of the thin film surface (cf.~Fig.~\ref{fig:fig2}). The RHEED intensity evolution shown in Fig.~\ref{fig:fig2} thus demonstrates that the film thickness can be controlled down to the individual IrO$_{2}$ sub-unit cell layer. In addition to SIO thin films on STO and NGO substrates, we have also investigated a bulk SIO single crystal sample grown by flux growth method as a reference.   

\subsection{Epitaxial strain of the Sr$_2$IrO$_4$ thin films}
\label{sec:EpiStrain}

\begin{figure}
\centering
\includegraphics[width=\columnwidth]{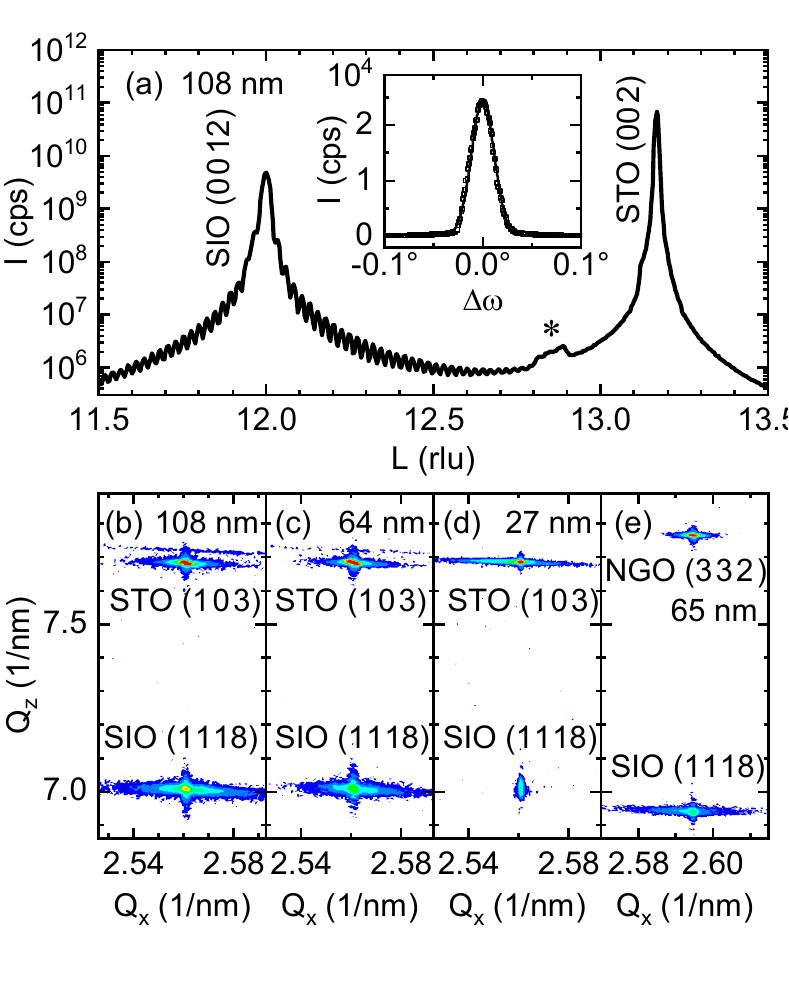}
\caption{(a) X-ray diffraction of a 108\,nm thick SIO film on a (001)-oriented STO substrate. The $L$-scan reveals strong Laue oscillations indicating a coherent growth over the whole thickness. The inset shows the rocking curve around the SIO (0\,0\,12) reflection. A full width at half maximum of $0.03^{\circ}$ is obtained by the Gaussian fit to the data (black line), demonstrating the high crystalline quality of the SIO film. The small powder reflections marked by the asterisks is caused by the beryllium dome used during the measurement.  (b)-(d) Reciprocal space maps of SIO thin films around the STO~(1\,0\,3) reflection and (e) around the NGO~(3\,3\,2) reflection, indicating that all SIO thin films are nearly fully strained.}  
\label{fig:fig3}
\end{figure}

To investigate the crystalline structure and the epitaxial relation of the SIO thin films with respect to the respective substrates, we performed detailed X-ray diffraction (XRD) measurements. A typical XRD $L$-scan around the SIO~(0\,0\,12) reflection is shown in Fig.~\ref{fig:fig3}(a). In all thin films, we found strong and pronounced Laue oscillations, indicating a coherent growth throughout the entire thickness of the thin films. The full width at half maximum (FWHM) of the rocking curves around the SIO~(0\,0\,12) reflection of less than $0.03^\circ$ reveal a low crystalline mosaic spread and an excellent crystalline quality of all SIO thin films (cf.~inset of Fig.~\ref{fig:fig3}(a)). The epitaxial relation between film and substrate was investigated by performing XRD reciprocal space maps (RSMs) using a conventional Bruker laboratory X-ray source. Typical RSMs around the STO~(1\,0\,3) and NGO~(3\,3\,2), respectively, are shown in Fig.~\ref{fig:fig3}(b)-(e). Additionally, RSMs were also performed around the STO~(1\,1\,3) reflection (not shown here). In both type of RSM measurements, the respective SIO reflections appear almost at the same in-plane $Q_x$ position as the respective STO and NGO substrate reflections. This result demonstrates that even the 108\,nm thick SIO film is still strained, exhibiting almost the same in-plane lattice parameter as the STO substrate. 

\begin{table}
\begin{ruledtabular}
\caption{Results of detailed XRD analysis of the SIO thin films with different thickness $d$ on STO and NGO substrates (sub.) using synchrotron light. The in-plane ($\epsilon_{xx}$) as well as out-of-plane ($\epsilon_{zz}$) strain state of the thin films are calculated from the measured lattice parameter ($a$, $c$) using the bulk SIO lattice parameter of Ref.~\onlinecite{Bhatti:2014} and assuming a tetragonal symmetry. For the calculation of the expected change of the tetragonality $\Delta d_\bot/d_\parallel$ and the rotation angle of the IrO$_6$ octahedra $\Delta\rho$ with respect to the unstrained SIO structure, the calculated variation of $d_\bot/d_\parallel$ and $\rho$ as a function of strain reported in Ref.~\onlinecite{Bhandari:2018} is used.}
\label{tab:table1}
\begin{tabular}{|c c c c c c c c c|} 
d (nm) & sub.  & $a$\,(\AA) & $c$\,(\AA) & $\epsilon_{xx}$\,(\%) & $\epsilon_{zz}\,(\%)$ & $\Delta d_\bot/d_\parallel$ & $\Delta\rho$ \\ \hline
27 &  STO &5.51 & 25.69 & 0.38 & -0.36 & -0.005 & -0.41$^\circ$\\
64 & STO & 5.51 & 25.67 & 0.40 & -0.24 & -0.006 & -0.48$^\circ$\\
79 & STO & 5.52 & 25.67 & 0.49 & -0.43 & -0.006 & -0.50$^\circ$\\
96\footnote{This SIO thin film was fabricated by using a non-stoichiometric target with an Ir-excess of around 30\%.} & STO & 5.50 & 25.72 & 0.29 & -0.24 & -0.003 & -0.33$^\circ$\\
108 & STO & 5.51 & 25.67 & 0.35 & -0.43 & -0.004 & -0.38$^\circ$\\\hline
 65 & NGO & 5.47 & 25.93 & -0.13 & 0.54 & 0.003 & +0.03$^\circ$\\
\end{tabular}
\end{ruledtabular}
\end{table}

From careful XRD orientation matrix refinements using synchrotron radiation and four-circle diffractometer setups at the I16 (Diamond Light Source) and XMaS (European Synchrotron Radiation Facility) beamlines, we have made an accurate measurement of the crystalline lattice parameters of our SIO thin films and calculated the strain state of the SIO thin films using the SIO bulk lattice parameters reported in Ref.~\onlinecite{Bhatti:2014}. The in-plane ($\epsilon_{xx}$) and out-of-plane ($\epsilon_{zz}$) strain of the thin films are listed in Table~\ref{tab:table1}. All SIO thin films on STO substrates exhibit an epitaxy-induced in-plane tensile strain between 0.35\% and 0.49\% nearly independent of the film thickness, except for the SIO thin film with a thickness of 96\,nm, which was fabricated by using a non-stoichiometric target with an Ir-excess of around 30\%. The difference of the in-plane strain state is most likely caused by a different Ir and O content in the respective SIO thin films,\cite{Llorente:2018} which might affect the magnetic properties.\cite{Kim:2017a} However, Table~\ref{tab:table1} shows that the 108\,nm thick SIO thin film is almost fully strained, which is in contradiction to the data reported by Serrao and coworkers.\cite{Serrao:2013,Sung:2016} They found a reduction of the in-plane strain state from 0.31\% to 0.17\% while increasing the film thickness from 5\,nm to 60\,nm. However, our results are somewhat in agreement with other groups,\cite{Afanasiev:2019,Lupascu:2014,Miao:2014,Nichols:2013} demonstrating that strained, epitaxial SIO thin films on STO substrates can be achieved up to a thickness of 108\,nm. In contrast to the SIO thin films on STO substrates, a compressive in-plane strain of -0.13\% was found for the 65\,nm thick SIO layer on the (110)-oriented NGO substrate. With the derived values of the in-plane strain $\epsilon_{xx}$, the change of the tetragonality $\Delta d_\bot/d_\parallel$ as well as the rotation angle of the IrO$_6$ octahedra $\Delta\rho$ with respect to the unstrained SIO structure was estimated using the strain-dependence of $d_\bot/d_\parallel$ and $\rho$ calculated by density-functional methods by Bhandari and coworkers.\cite{Bhandari:2018} As obvious from Table~\ref{tab:table1}, the differences with respect to the unstrained bulk structure are small, indicating that we expect only marginal changes of the SIO bulk properties.

\subsection{Rotation angle of the oxygen octahedra of the Sr$_2$IrO$_4$ thin films}
\label{sec:RotOcta}

The epitaxial strain of the SIO thin films should lead not only to changes of the tetragonality discernable by different lattice constants with respect to the SIO bulk material but also to modifications of the rotation angle of the IrO$_6$ octahedra $\Delta \rho$ and therefore of the Ir-O-Ir bond angle (see Fig.~\ref{fig:fig1}). In Table~\ref{tab:table1}, we estimated small changes of $\rho$ from the unstrained bulk value $\rho=11.8^\circ$ based on the density-functional calculations by Bhandari and coworkers.\cite{Bhandari:2018} In order to experimentally verify these values, we performed detailed X-ray scattering experiments on the tensile strained 108\,nm thick SIO thin film grown on STO and the compressively strained 65\,nm SIO film on NGO as well as on the SIO bulk single crystal for reference. In unstrained bulk Sr$_2$IrO$_4$, the rotation of the oxygen octahedra schematically shown in Fig.~\ref{fig:fig1}(b), gives rise to additional Bragg peaks characterized by the general reflection condition $H+K=\mathrm{odd}$ and $L=\mathrm{odd}$.\cite{Crawford:1994} For the SIO bulk single crystal, we therefore examined the (1\,2\,25) Bragg reflection, which fulfills the above condition.   

\begin{figure}
	\centering
	\includegraphics[width=\columnwidth]{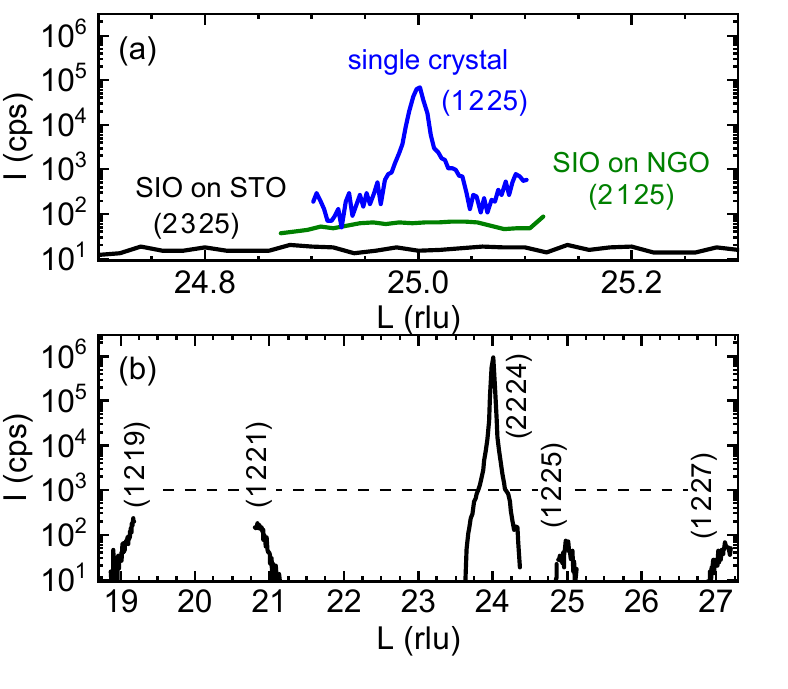}
	\caption{$L$-scans taken at several Bragg reflections corresponding to the rotation of the oxygen octahedra with conditions $H+K=\mathrm{odd}$ and $L=\mathrm{odd}$.\cite{Crawford:1994} (a) $L$-scans around $L=25$ of the SIO single crystal (blue line), the compressively strained 64\,nm thick SIO thin film on NGO (green line), and the tensile strained 108\,nm thick SIO thin film on STO (black line) recorded at the (1\,2\,25), (2\,1\,25), and (2\,3\,25) Bragg reflections, respectively. (b)	(1\,2\,$L$)-reflections of the 108\,nm thick SIO thin film on STO. A broad background slope is visible for the (1\,2\,19) and (1\,2\,21) reflections and a very weak intensity is observed for the (1\,2\,25) and (1\,2\,27) reflections. This is well below the intensity calculated for these reflections (dashed vertical line) by assuming the intensity ratio between these reflections and the structural (2\,2\,4) reflection of the SIO thin film to be the same as that of the SIO bulk sample.} 
	\label{fig:fig4}
\end{figure}  

As shown in Fig.~\ref{fig:fig4}(a), the (1\,2\,25) Bragg reflection of the bulk SIO sample (blue line) is very sharp and corresponds to a long-range correlated order of the rotation of the oxygen oxtahedra along the crystallographic $c$-direction of the lattice. The intensity of this reflection is around three orders of magnitude weaker than the structural reflections such as the (0\,0\,24) Bragg reflection. By using the intensity of the off-specular structural Bragg reflection (2\,2\,24) of the 108\,nm thick SIO thin film, which is found to be of the order of $10^{6}$\,counts/sec (see Fig.~\ref{fig:fig4}(b)), we can therefore estimate the intensity for the Bragg peaks corresponding to the rotation of the oxygen octahedra in the 108\,nm thick SIO thin film on STO (cf.~horizontal black dashed line in Fig.~\ref{fig:fig4}(b)). However, as obvious from Fig.~\ref{fig:fig4}, we do not find clear evidence of a finite oxygen octahedra rotation structure in our SIO thin films. For the investigated Bragg reflections (1\,2\,19), (1\,2\,21), (1\,2\,25), (1\,2\,27), and (2\,3\,25) of the tensile strained 108\,nm thick SIO thin film, some of the reflections show a sloping background or small peaks well below the expected intensity (see Fig.~\ref{fig:fig4}(a),(b)). Furthermore, we also do not find any significant diffraction intensity at reflections corresponding to rotation of the oxygen octahedra from the compressively strained 65\,nm thick SIO thin film on NGO (cf.~green line in Fig.~\ref{fig:fig4}(a)). One possibility to explain the data is that these Bragg peaks are very broad and barely detectable, rather than a complete absence of a rotation of the oxygen octahedra in our SIO thin films as found in the Ba$_2$IrO$_4$ compound\cite{Okabe:2011} as well as in strained SrIrO$_{3}$ thin films.\cite{Guo:2020} This interpretation is further supported by the observation of a finite net magnetization of our SIO thin films under applied magnetic fields (see Fig.~\ref{fig:fig5}), which we presume to arise from canting of the Ir$^{4+}$-magnetic moments. Together with the assumption of an almost complete locking of the canting of the moments to the rotation of the oxygen octahedra as observed in the bulk SIO compound, a finite rotation of the oxygen octahedra should be present in the SIO thin films. We therefore propose that the weak scattering shown in Fig.~\ref{fig:fig4} corresponds to a very short correlation length, as also reported by Lupascu and coworkers.\cite{Lupascu:2014} This would correspond to rotations of the oxygen octahedra that are not coherent along the crystallographic $c$-axis in our SIO thin films. 

Another possible explanation for the observed weak intensity is that the stacking sequence of the octahedra is different in all investigated SIO thin films compared to that of the bulk oxygen ocathedra tilting structure. However, our data so far can not provide any further information on this possibility. 

The conclusion drawn from our investigation of the rotation of the oxygen octahedra is that the structure of the oxygen octrahedra in our thin films is clearly different from the bulk compound. This difference appears to be present regardless of thickness or strain state (tensile or compressive) of the SIO films. Our experimental results clearly show that the long-range order of the rotation of the oxygen octahedra is not re-established by some structural relaxation mechanism when increasing the film thickness up to 108\,nm. We thus conclude that the SIO thin films presented here are strained up to a thickness of 108\,nm, with no sign of either external (lattice constants) or internal (oxygen rotation angle) strain relaxation.

\subsection{Integral magnetic properties of the Sr$_2$IrO$_4$ thin films}
\label{sec:IntMagProp}

\begin{figure}
\centering
\includegraphics[width=\columnwidth]{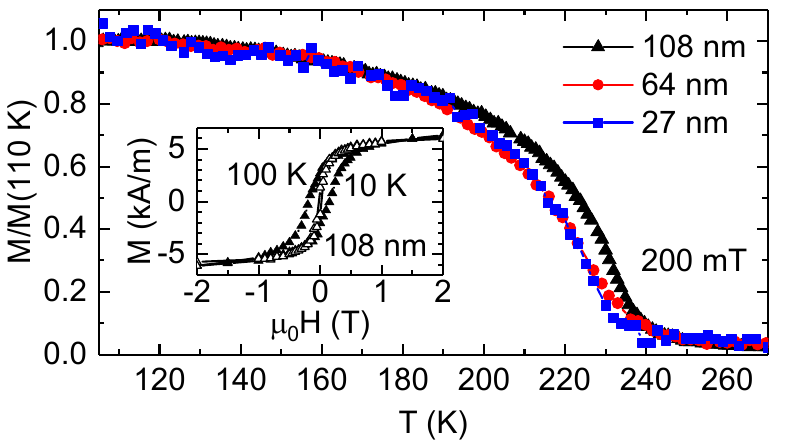}
\caption{Normalized magnetization $M/M(110\,\mathrm{K})$ along the [110]-direction of SIO as a function of temperature of SIO thin films with different thicknesses on STO substrates. The data is measured with an in-plane magnetic field of 200\,mT, which is above the metamagnetic phase transition of SIO, by SQUID magnetometry. The data were recorded while heating the samples from 10\,K to 300\,K after cooling in a magnetic field of 7\,T (field-cooling). The magnetization is normalized to that at 110\,K, which is above the structural phase transitions of the STO substrate. Inset: Magnetization versus magnetic field of the 108\,nm thick SIO thin film measured at 10\,K (full symbols) and 100\,K (open symbols). The linear diamagnetic background of the STO substrate is subtracted by assuming a saturation of the magnetization for magnetic fields $\mu_0 H > 2$\,T.}
\label{fig:fig5}
\end{figure}

The normalized, integral magnetization $M$ measured along the [110] in-plane direction of SIO as a function of temperature $T$ of three SIO thin films with different thicknesses deposited on STO substrates is shown in Fig.~\ref{fig:fig5}. The measurements were carried out using a SQUID magnetometer with an in-plane magnetic field of 200\,mT, i.e. above the meta-magnetic transition of SIO,\cite{Kim:2009} while heating the respective sample from 10\,K to 300\,K after field-cooling in a magnetic field of 7\,T. The $M(T)$-curves reveal a N\'eel temperature $T_N$ of 240\,K of the 108\,nm thick SIO thin film, which is almost identical to the bulk value.\cite{Crawford:1994,Shimura:1995,Cao:1998} While reducing the thin film thickness, $T_N$ decreases to 238\,K and 232\,K for the 64\,nm and 27\,nm SIO thin film, respectively. Therefore, Fig.~\ref{fig:fig5} reveals that $T_N$ is slightly reduced upon decreasing the thin film thickness. However, in contrast to the data reported by Lupascu and coworkers,\cite{Lupascu:2014} we find no indication of an enhancement of the magnetic transition temperature with strain, which is in agreement with other reports on SIO thin films.\cite{Miao:2014,Lu:2015,Seo:2019}

Furthermore, magnetic hysteresis loops $M(H)$ of the 108\,nm thick SIO thin film on STO reveal a saturation magnetization of 6.3\,kA/m and 6.1\,kA/m, which corresponds to an effective magnetization of $0.53\,\mu_\mathrm{B}/\mathrm{f.u.}$ and $0.51\,\mu_\mathrm{B}/\mathrm{f.u.}$, measured at 10\,K and 100\,K, respectively (cf.~inset in Fig.~\ref{fig:fig5}). These values are in good agreement with measurements on polycrystalline and single-crystalline bulk samples\cite{Crawford:1994,Ge:2011,Bhatti:2014} and demonstrate that the tensile in-plane strain of the SIO thin films on STO substrates of 0.38-0.45\% only slightly change the canting of the Ir$^{4+}$-moments and therefore the net magnetic moments.\cite{Kim:2016a,Bhandari:2018} As a consequence, this indicates that the angle of the rotation of the oxygen octahedra discussed in the previous section might indeed be similar to the bulk value assuming equal magnetoelastic coupling. However, a finite remanent magnetization is not expected in bulk SIO, since the net magnetic moments within the individual IrO$_2$ layer are expected to cancel each other in the $uddu$ antiferromagnetic out-of-plane stacking sequence at zero magnetic field. The finite remanent magnetization and magnetic hysteresis at low temperature could be caused by the elastic clamping of the SIO thin film onto the STO substrate, which undergoes structural phase transitions for $T < 100$\,K.\cite{Shirane:1969} These structural transitions are accompanied by finite rotations of the oxygen octahedra or even anti-phase domains in surface-near regions of STO,\cite{Loetzsch:2010} leading to the formation of structural domains, which might pin the ferromagnetic $uuuu$-stacking sequence of SIO. However, Sung and coworkers\cite{Sung:2016} reported that the magnetic properties in single crystalline SIO are strongly dependent on the oxygen stoichiometry. Therefore, the finite magnetic hysteresis found in our SIO thin films might point to a finite density of oxygen vacancies within the SIO lattice.

\section{Magnetic $A$-sublattice}
\label{sec:A-sublattice}

\begin{figure}
\centering
\includegraphics[width=0.8\columnwidth]{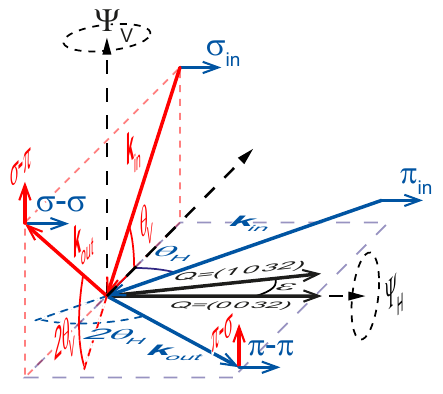}
\caption{Schematic of the polarized X-ray resonant scattering (XRS) geometry employed in our experiments using vertical (purple dashed lines) and horizontal (light blue dashed lines) scattering planes. The red (blue) arrows show the incoming and scattered X-rays with wave vector $\mathbf{k}_\mathrm{in}$ and $\mathbf{k}_\mathrm{out}$ as well as angles $\theta_\mathrm{V}$ and $2\theta_\mathrm{V}$ ($\theta_\mathrm{H}$ and $2\theta_\mathrm{H}$) in the vertical (horizontal) scattering plane, respectively. Polarization analysis has been used to determine the polarization of the scattered X-rays with respect to that of the incoming X-rays ($\sigma$-$\sigma$, $\sigma$-$\pi$, $\pi$-$\sigma$, $\pi$-$\pi$). The azimuthal dependence of the XRS was probed around $\Psi_\mathrm{H}$ and $\Psi_\mathrm{V}$. In addition, the $Q$ vector of the (1\,0\,32) reflection, which is tilted by an angle $\epsilon$ from $Q=(0\,0\,31)$, is shown.}  
\label{fig:fig6}
\end{figure}

To investigate the basal-plane antiferromagnetic $A$-sublattice of our SIO thin films, we have employed polarized X-ray resonant scattering (XRS) measurements at the Ir $L_{2,3}$ absorption edges. The measurements were carried out at the XMaS and D2AM bending magnet beamlines of the European Synchrotron Radiation Facility (ESRF) as well as at the I16 insertion device beamline of the Diamond Light Source. At XMaS and D2AM, our XRS studies were performed in the vertical scattering plane (shown by red lines in Fig.~\ref{fig:fig6}) with incident $\sigma_\mathrm{in}$-polarized photons, which is defined by the electric vector perpendicular to the scattering plane. Polarization analysis was used to determine the polarization of the scattered beam: $\sigma$-$\sigma$ (same polarization as the incoming X-rays $\sigma_\mathrm{in}$), $\sigma$-$\pi$ (rotated polarization with respect to $\sigma_\mathrm{in}$). At XMaS and D2AM,\cite{CRG-D2AM} the incident flux is around $5\times10^{11}$ photons per second and the resolution $\Delta E/E=\Delta \lambda/\lambda \sim 10^{-4}$. At I16, XRS experiments were additionally carried out in the horizontal scattering plane (shown by blue lines in Fig.~\ref{fig:fig6}) with incident $\pi_\mathrm{in}$-polarized photons, defined with their electric vector in the scattering plane. At I16, the incident flux is around $1\times10^{13}$ photons per second and the resolution $\Delta E/E=\Delta\lambda/\lambda\sim10^{-4}$. In both experiments, azimuthal scans were performed, where the sample is rotated about the scattering vector $\mathbf{Q}$ with the azimuthal angles $\Psi_{V}$ and $\Psi_{H}$, respectively.

In such a diffraction experiment, the antiferromagnetic $A$ sublattice gives rise to magnetic Bragg reflections below the N\'eel temperature $T_N$, indexed by $(1\,0\,4n+2)$ and $(0\,1\,4n)$\cite{Kim:2009,Boseggia:2013} for a $uddu$-stacking sequence.\cite{Porras:2019} As the tetragonal structure of SIO allows for a twinned domain structure\cite{Dhital:2013} with $uddu$- and $uudd$-stacking sequences,\cite{Porras:2019} we expect magnetic reflections indexed by $(1\,0\,4n+2)$ and $(0\,1\,4n)$ as well as $(1\,0\,4n)$ and $(0\,1\,4n+2)$, i.e., at all even $L$-positions of $(1\,0\,L)$ and $(0\,1\,L)$.\cite{Kim:2009,Lupascu:2014}  

\subsection{$J_\mathrm{eff}=1/2$ ground state in Sr$_2$IrO$_4$ thin films}

\begin{figure}
\centering
\includegraphics[width=\columnwidth]{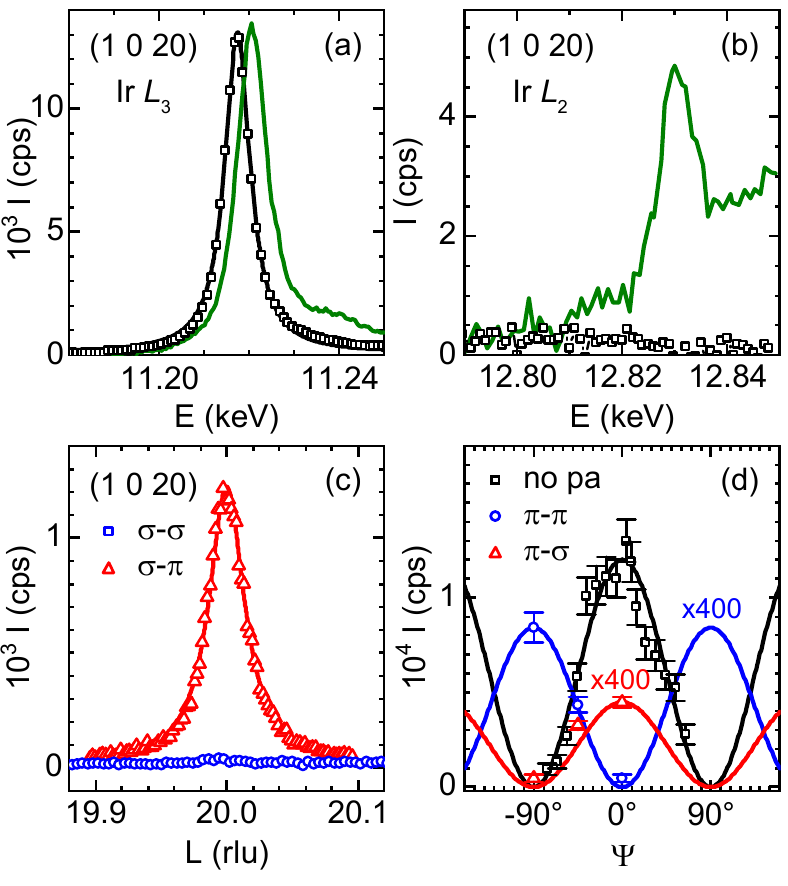}
\caption{XRS measurements at $T=20$\,K of the 108\,nm thick SIO thin film on STO: (a), (b) Energy scans around the Ir $L_{3}$-edge and Ir $L_{2}$-edge with fixed wave-vector $\mathbf{Q}$ of the magnetic reflection $(1\,0\,20)$ (black open symbols). The fluorescence is shown by the green line. (c) Typical $L$-scan at the Ir $L_{3}$-edge around the $(1\,0\,20)$ magnetic reflection with polarization analysis $\sigma$-$\sigma$ (blue open symbols) and $\sigma$-$\pi$ (red open symbols), demonstrating that the scattering is all in the $\sigma$-$\pi$ channel. (d) The azimuthal $\Psi$-dependence of the $(1\,0\,20)$ reflection at the Ir $L_{3}$-edge without polarization analysis (no pa, black open symbols) and the $(1\,0\,32)$ reflection taken with polarization analysis and with $\pi$ incident polarized photons (blue and red open symbols). The data of the XRS measurements around the $(1\,0\,32)$ reflection are multiplied by 400. $\Psi=0^\circ$ corresponds to the $[1\,0\,0]$ direction. The solid lines are fits to the data of the azimuthal dependence using the bulk antiferromagnetic structure with magnetic moments pointing along the $a$-axis.} 
\label{fig:fig7}

\end{figure}

The result of XRS measurements on the 108\,nm thick SIO thin film on STO at the magnetic sublattice reflections (1\,0\,20) and (1\,0\,32) are shown in Fig.~\ref{fig:fig7}. 
The energy scans around the Ir $L_{3}$-edge at fixed wave-vector $\mathbf{Q}$ corresponding to the $(1\,0\,20)$ magnetic reflection shows a strong enhancement of the XRS signal (black open symbols in Fig.~\ref{fig:fig7}(a)), reaching a maximum value at $E=11.217$\,keV, which is approximately 3\,eV below the maximum of the Ir $L_{3}$-fluorescence (green line in Fig.~\ref{fig:fig7}(a)). The energy separation between the maximum XRS signal and the maximum in fluorescence is in agreement with the crystal field splitting of 3\,eV of the $t_{2g}$ and $e_{g}$ states of SIO.\cite{Katukuri:2012,Sala:2014} A fit to the data of the energy scan using a Lorentzian peak function (black line in Fig.~\ref{fig:fig7}(a)) reveals a full width at half maximum (FWHM) of 6.8\,eV, similar to those found for bulk single crystals.\cite{Kim:2009,Boseggia:2013} In contrast to the Ir $L_{3}$-edge, no XRS enhancement of the scattered intensity is observed at the Ir $L_{2}$ absorption edge (see Fig.~\ref{fig:fig7}(b)). The suppressed XRS at this edge is consistent with previous XRS experiments on bulk SIO single crystals\cite{Kim:2009} and other iridate compounds.\cite{Kim:2012b,Calder:2012,Ohgushi:2013,Boseggia:2013a} Further insight into the strong XRS signal at the Ir $L_{3}$-edge can be gained by employing polarization analysis of the scattered signal. A typical $L$-scan around the (1\,0\,20) magnetic reflection with polarization analysis of the scattered beam into unrotated $\sigma$-$\sigma$ (blue open symbols) and rotated $\sigma$-$\pi$ (red open symbols) components is shown in Fig.~\ref{fig:fig7}(c) evidencing that all the scattering is in the $\sigma$-$\pi$ channel. This is consistent with electric dipole ($E1$) transitions from the core $2p$ orbitals to the $5d$ polarized states.\cite{Matteo:2012,Boseggia:2013} 

The vanishing magnetic XRS intensity at the Ir $L_{2}$-edge together with a finite intensity at the Ir $L_{3}$-edge in the cross-polarized channel have been proposed as a characteristic fingerprint of the $J_\mathrm{eff}$=$\frac{1}{2}$ ground state.\cite{Kim:2009} However, some doubts have been raised, as to whether this signature is a definite proof.\cite{Chapon:2011,Haskel:2012} In particular, Moretti-Sala and coworkers have shown that this is only true, if the Ir$^{4+}$-moments lie within the $ab$-plane.\cite{Sala:2014a} Azimuthal scans, i.e., the rotation of the sample about the magnetic scattering vector $Q$ with respect to the incident and scattered polarizations can provide direct information on the orientation of the component of the ordered magnetic moment measured with XRS.\cite{Chapon:2011} The azimuthal dependence of the (1\,0\,20) magnetic reflection with incident $\sigma$-polarized photons and without polarization analysis is shown in Fig.~\ref{fig:fig7}(d) (black open symbols) together with the azimuthal dependence of the (1\,0\,32) magnetic reflection with incident $\pi$-polarized photons and with polarization analysis of the scattered beam into $\pi$-$\pi$ (blue open symbols) and  $\pi$-$\sigma$ (red open symbols) channels. The solid lines in Fig.~\ref{fig:fig7}(d) are fits to the data of the azimuthal dependence using the bulk antiferromagnetic structure with magnetic moments pointing along the $a$-axis. Therefore, the measured azimuthal dependence in Fig.~\ref{fig:fig7}(d) is very similar to that of the bulk SIO compound, with magnetic moments in the $ab$-plane and directed along the $a$-axis. The XRS results shown in Fig.~\ref{fig:fig7} thus reveal that while the 108\,nm thick SIO thin film appears to be fully strained (cf.~Fig.~\ref{fig:fig3} and Table~\ref{tab:table1}), the crystal field splitting, the magnetic $A$-sublattice structure as well as the $J_\mathrm{eff} = \frac{1}{2}$ ground state is very similar to that of the bulk SIO compound.

\subsection{Temperature dependence of the magnetic $A$-sublattice}
\label{sec:T-depA}

\begin{figure}
\centering
\includegraphics[width=\columnwidth]{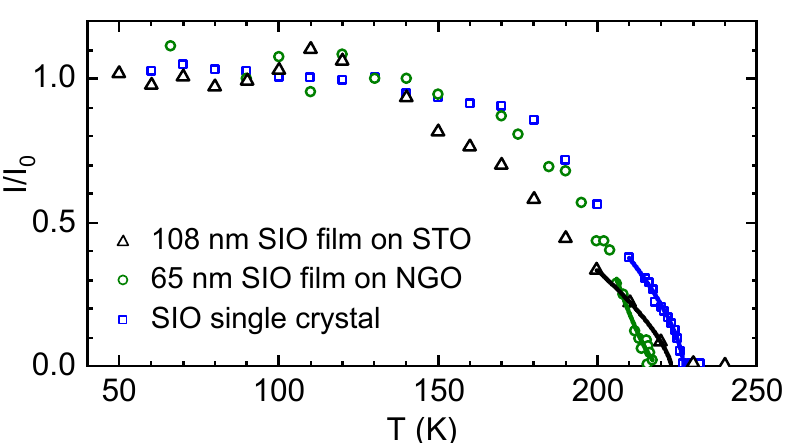}
\caption{Normalized integrated intensity from $L$-scans around the (1\,0\,20) magnetic reflection of the bulk SIO single crystal (blue symbols), the 108\,nm SIO film grown with tensile strain on a STO substrate (black symbols) and the 65\,nm SIO thin film grown under compressive strain on a NGO substrate (green symbols). The solid lines are power law fits to the data using Eq.~(\ref{eq:IT}).}
\label{fig:fig8}
\end{figure}

The relatively strong magnetic scattering in our SIO thin films has enabled us to undertake a detailed analysis of the thermal evolution of the XRS signal related to the magnetic $A$-sublattice.\cite{Lupascu:2014} The normalized integrated intensities derived from $L$-scans around the (1\,0\,20) magnetic Bragg reflection for the tensile strained 108\,nm thick SIO film on STO, the compressive strained 65\,nm thick SIO film on NGO, and the SIO single crystal are shown in Fig.~\ref{fig:fig8}. The temperature dependencies of the intensity near the N\'eel temperature $T_N$ have been modelled by a standard power-law expression
\begin{equation}
  I\propto \left[1-\frac{T}{T_{N}}\right]^{2\beta} 
	\; ,
	\label{eq:IT}
\end{equation}
where $\beta$ denotes the critical exponent of the phase transition.\cite{Stanley:1971} For the SIO single crystal (blue open symbols), we find a critical exponent of $\beta=0.19$ and a N\'eel temperature $T_{N}=227$\,K in excellent agreement with recent XRS and neutron scattering studies.\cite{Ye:2013,Dhital:2013,Vale:2015} Therefore, $\beta$ deviates significantly from the values expected for a 2D Ising model ($\beta=0.125$), the pure 2D $XY$ model ($\beta=0.23$),\cite{Bramwell:1993} and a 3D Heisenberg system ($\beta\sim0.35$), but is consistent with the value for the 2D $XYh_4$ universality class including an additional fourfold anisotropy term $h_4$ in the Hamiltonian.\cite{Taroni:2008} Since $\beta$ varies slowly with $h_4$, the strength of this additional anisotropy can be estimated, which results in $h_4=0.61$, indicating that the additional in-plane anisotropy is important in SIO. This supports the recent findings by Vale and coworkers\cite{Ye:2013} demonstrating that the critical fluctuations of SIO can be described within a two-dimensional (2D) anisotropic Heisenberg model, where the main Heisenberg interactions are augmented by the 2D $XY$ anisotropy in SIO. 
For our strained SIO thin films, we find $T_{N}=223$\,K for the tensile strained 108\,nm film and $T_{N}=219$\,K for the compressive strained 65\,nm SIO thin film, respectively. As obvious from Fig.~\ref{fig:fig8}, we do not find an enhancement of $T_N$ with strain in our SIO thin films. This is in contrast to the data reported by Lupascu and coworkers,\cite{Lupascu:2014} but in agreement with our SQUID magnetometry measurements shown in Fig.~\ref{fig:fig5}. The observed difference in $T_{N}$ of the 108\,nm thick SIO film on STO compared to the integral magnetization measurements by SQUID magnetometry, which is sensitive to the net magnetic moment caused by the canting of the Ir$^{4+}$-moments, is mainly caused by the different applied magnetic field. SQUID magnetometry measurements at 0\,T reveal a transition temperature of 229\,K (not shown here). Another possibility might be weak inhomogeneities resulting in regions with slightly different transition temperatures inside the SIO thin films. However, taking the sharpness of the magnetic transition, which is similar to that of the SIO single crystal, into account, this seems to be only a minor effect. Unfortunately, the less data points around $T_N$ of the SIO thin films compared to the SIO single crystal does not allow a clear determination of the critical exponent $\beta$ of the SIO thin films. We find $\beta=0.36 \pm 0.1$ depending on the fitting range. However, it seems that $\beta$ of the SIO thin films is larger compared to the bulk value, pushing the critical properties of SIO away from a $2D$ towards a more $3D$ Heisenberg model.

\subsection{Thickness dependence of the magnetic correlation length}

\begin{figure}
\centering
\includegraphics[width=\columnwidth]{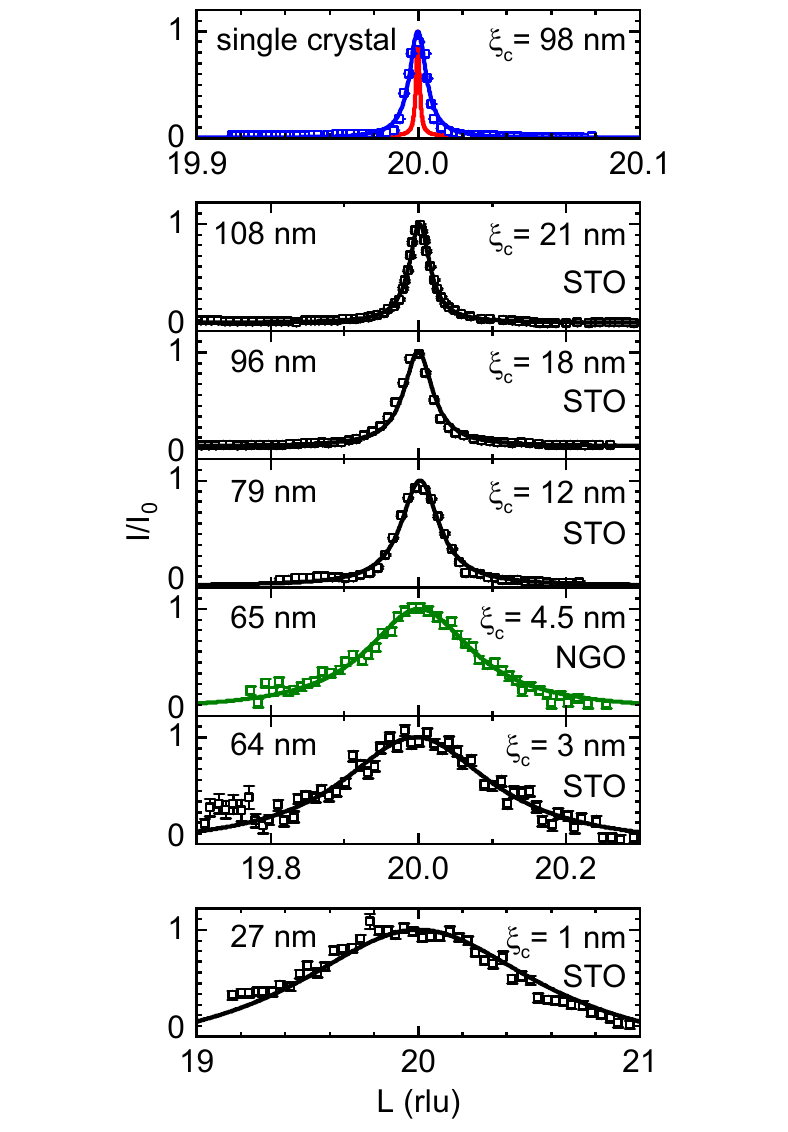}
\caption{$L$-scans around the (1\,0\,20) magnetic reflection used to determine the $c$-axis magnetic correlation length $\xi_c$ of (from top to bottom) the SIO bulk single crystal, a 108\,nm SIO film on STO, a 96\,nm SIO film on STO, a 79\,nm film on STO, a 65\,nm SIO film on NGO, a 64\,nm film on STO, a 27\,nm film on STO. Note the different $x$-axis scales. The measurements were carried out at $T=150$\,K, well below the N\'eel temperature of SIO. The $c$-axis magnetic correlation length $\xi_c$ is extracted from Lorentzian fits to the data (solid lines). The resolution of the experimental set-up is shown by the red line.}
\label{fig:fig9}
\end{figure}

To further examine the antiferromagnetic properties of the SIO thin films, we investigate the magnetic correlation length along the $c$-axis $\xi_{c}$ of SIO. To this end, we performed $L$-scans around the (1\,0\,20) magnetic Bragg reflection (without polarization analysis and with incident $\sigma$-polarized photons) and extracted $\xi_{c}$ from the half width of  Lorentzian fits to the data. As a reference, we first discuss the result of the XRS of the SIO single crystal. As obvious from the blue symbols in Fig.~\ref{fig:fig9}, the $L$-scan around the (1\,0\,20) magnetic Bragg reflection is very sharp, demonstrating a long-range magnetic order along the $c$-axis with a correlation length of $\xi_{c}=98\,\mathrm{nm}\approx 153\,c_0$ with $c=4\,c_0$.\cite{Fujiyama:2012} Furthermore, we found a long-range magnetic order within the $ab$-plane from $H$-scans around (1\,0\,20) (not shown here). From these measurements, we extracted a magnetic correlation length of $\xi_{ab}=109\,\mathrm{nm}\approx 280\,a_0$ with the average Ir-Ir nearest neighbor distance $a_0\approx 0.39$\,nm, demonstrating a strong in-plane exchange coupling with $\xi_{ab} \gg a_0$ and out-of plane correlation length with $\xi_{c} > c_0$ in bulk SIO at 150\,K. These values are comparable to recent XRS-experiments.\cite{Fujiyama:2012} 

For our strained SIO thin films, Fig.~\ref{fig:fig9} reveals first the absence of thickness fringes around the (1\,0\,20) magnetic reflection, which is consistent with a magnetic order that is not fully coherent through the entire film thickness. Second, the $L$-scans disclose a strong dependence of the magnetic correlation length $\xi_c$ along the $c$-axis on the SIO thin film thickness. We find a reduced correlation length of $\xi_c=21$\,nm for the 108\,nm thick SIO film on STO compared to the SIO bulk single crystal. The $\xi_c$ value is further decreased dramatically to $\xi_c=1$\,nm for the 27\,nm thick SIO thin film on STO. Interestingly, the compressive strained, 65\,nm thick SIO film on NGO exhibits a similar $\xi_c$ value as the tensile strained SIO thin film on STO with almost the same thickness (cf.~green symbols in Fig.~\ref{fig:fig9}).

\begin{figure}
\centering
\includegraphics[width=\columnwidth]{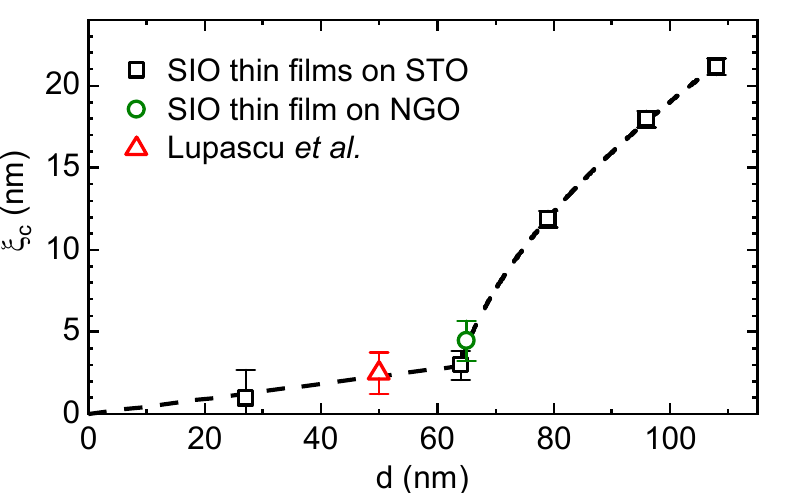}
\caption{Out-of-plane magnetic correlation length $\xi_c$ as a function of SIO thin film thickness $d$ deduced from Lorentzian fits to $L$-scans around the (1\,0\,20) magnetic Bragg reflection (see Fig.~\ref{fig:fig9}). The black symbols denote $\xi_{c}$ of tensile strained SIO thin films on STO, while the green symbol marks $\xi_{c}$ of the compressive strained SIO thin film on NGO. The $\xi_c$-value of an epitaxial SIO thin film on STO reported in Ref.~\onlinecite{Lupascu:2014} is also included in the figure (red symbol). The dashed black line is a power-law fit, pointing to a large growth in $\xi_c$ occurs above a critical thickness of $d_{c}=62.5$\,nm.}
\label{fig:fig10}
\end{figure}

The extracted $\xi_c$ values as a function of thickness $d$ is shown in more detail in Fig.~\ref{fig:fig10}. From this figure, two regimes can be clearly identified: A first regime for thickness $d$ below around 64\,nm with magnetic correlation lengths $\xi_{c}$ of similar scale, and a second regime for $d \gtrsim 65$\,nm, where the magnetic correlation length strongly increases. A power-law fit to the data reveals a critical thickness of $d_c=62.5$\,nm (cf. black dashed line in Fig.~\ref{fig:fig10}). This thickness dependence can not be ascribed to different strain states of the SIO thin films, since our SIO thin films are still strained up to a film thickness of 108\,nm (cf.~Table~\ref{tab:table1}). Furthermore, the correlation length for the compressive strained 65\,nm thick SIO thin film on NGO (green symbol in Fig.~\ref{fig:fig10}) agree well with the thickness dependence of the tensile strained SIO thin films on STO. Moreover, the $\xi_{c}$ value of a 50\,nm SIO thin film on STO reported in a previous XRS study by Lupascu \textit{et al.}\cite{Lupascu:2014} (red symbol in Fig.~\ref{fig:fig10}) fits perfectly to our findings. Therefore, our data strongly suggest that the magnetic correlation length along the $c$-axis $\xi_{c}$ is independent of the strain state of the SIO films, but highly thickness dependent with a critical thickness of around $d_c=62.5$\,nm separating regions with short ($\xi_{c} \lesssim c$) and long-range ($\xi_{c} > c$) magnetic order along the $c$-axis of SIO. We note that this critical thickness $d_c$ is not correlated to static regions of non-magnetic order, often referred to as magnetically dead layer, since static SQUID-magnetometry measurements reveal a clear magnetic signal also for SIO thin films with a thickness of 27\,nm (cf.~section~\ref{sec:IntMagProp}). Furthermore, we would then expect the correlation length to grow linearly with thickness above this value.  Therefore, the thickness-dependence of the correlation length $\xi_{c}$ might be more subtle, which we propose to involve changes in strongly correlated but highly fluctuating Ir$^{4+}$-moments with $d_c$ representing a transition from a quasi paramagnetic-like, fluctuating state towards a state with quasi-long-range order of the moments. This might also be related to the starting point of a reduced disorder in the oxygen lattice, although we could not observe a clear sign of a long-range ordering of the rotation of the oxygen octahedra in the 108\,nm thick SIO thin film (see section \ref{sec:RotOcta}). 

The conclusion drawn from our investigation of the out-of-plane correlation length $\xi_{c}$ is that the magnetic order can be dramatically tuned by varying the film thickness of SIO. The good agreement between the data and fit, together with the precise control of the thickness demonstrated in Fig.~\ref{fig:fig2}, indicates that film thickness can be used as a finely tuned dial to tune the out-of-plane magnetic correlation length $\xi_{c}$.

\section{Magnetic $B$-sublattice}
\label{sec:B-sublattice}

In contrast to the basal-plane antiferromagnetic $A$-sublattice, the magnetic $B$-sublattice arises from the canting of the Ir$^{4+}$-moments and thus depends strongly on the canting angle $\phi$ (cf. Fig.~\ref{fig:fig1}(b)-(d)). Therefore, the magnetic reflections (0\,0\,$2n+1$) linked to the magnetic $B$-sublattice will have zero intensity when there is no canting ($\phi=0^\circ$), as it is the case for Ba$_2$IrO$_4$,\cite{Boseggia:2013a} and finite intensity for $\phi\neq0^\circ$. An estimation of the canting angle $\phi$ can be obtained by determining the ratio between the intensity of the magnetic peaks sensitive to the $A$-sublattice ($(1\,0\,4n)$ and $(0\,1\,4n+2)$) and those for the $B$-sublattice (0\,0\,$2n+1$).\cite{Boseggia:2013} Despite the fairly strong XRS observed for the (1\,0\,$4n$)-type reflections (cf.~Fig.~\ref{fig:fig7}), we have been unsuccessful in observing any clear signal of the (0\,0\,$2n+1$)-type of magnetic reflections in our SIO thin films investigated. In particular, our searches with incident $\sigma$-polarized photons were hindered by a fairly high background from the specular charge scattering truncation rods and tails of the structural Bragg peaks of the films. The best data we could obtain on the canting of the magnetic structure was taken on the I16 undulator beamline using incident-$\pi$ polarized photons in the horizontal scattering geometry with polarization analysis of the scattered beam. In this geometry, the elastic charge scattering background is minimized by selecting magnetic peaks with a scattering angle $2\theta$ close to $90^{\circ}$ and the fluorescence background is suppressed by the polarization analyzer. With this set-up the background measured in the $\pi$-$\sigma$ channel was reduced to around 0.4\,counts/sec. 

\begin{figure}
\centering
\includegraphics[width=\columnwidth]{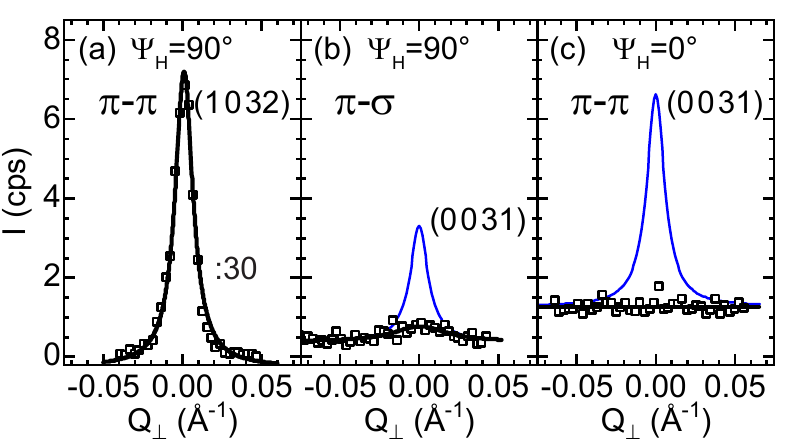}
\caption{(a)-(c) XRS of the 108\,nm thick SIO thin film on STO along the transverse magnetic wave-vector $Q_\perp$ in the horizontal scattering geometry with given azimuthal angles $\Psi_H$ around the (1\,0\,32) and (0\,0\,31) reflections using incident-$\pi$ polarized photons and polarization analysis of the scattered beam ($\pi$-$\pi$ and $\pi$-$\sigma$ channels). The intensity of the (1\,0\,32) reflection is divided by 30 for better visibility. The black lines display Lorentzian fits to the data. For comparison, the blue lines in (b) and (c) represent the expected intensities for a tilting angle of $\phi_\mathrm{bulk}$=12.2$^{\circ}$ of bulk SIO (see~Table~\ref{tab:table2}).}
\label{fig:fig11}
\end{figure}

The results of the XRS measurements around the (1\,0\,32) reflection related to the magnetic $A$-sublattice and the $B$-sublattice magnetic reflection (0\,0\,31) with an azimuth angle of $\Psi_H=90^\circ$ are shown in Fig.~\ref{fig:fig11}(a)-(c). As obvious from Fig.~\ref{fig:fig11}(a), a clear XRS signal is observed for the (1\,0\,32) reflection in the $\pi$-$\pi$ channel. This is consistent with the XRS results shown in Fig.~\ref{fig:fig7}(c) performed in the vertical scattering plane with $\Psi_V=0^\circ$, since the magnetic moments of the $A$-sublattice are perpendicular to the scattering plane for $\Psi_H=90^\circ$ and thus scatter into the $\pi$-$\pi$ channel. However, a weak intensity of the (0\,0\,31) reflection is only visible in the $\pi$-$\sigma$ channel (cf. Fig.~\ref{fig:fig11}(b)), while no (0\,0\,31) reflection could be observed in the unrotated $\pi$-$\pi$ polarization channel. This is mainly caused by the higher background signal in the $\pi$-$\pi$ channel compared to that of the $\pi$-$\sigma$ channel.  In order to analyze the magnetic scattering at the (1\,0\,32) and (0\,0\,31) magnetic reflections, we exploit the well known XRS cross-section for incident $\pi$-polarized light,\cite{Hill:1996} together with the structure factors for the (1\,0\,$4n$) and (0\,0\,$2n+1$) type reflections.\cite{Boseggia:2013} Thus, the expected intensities for the (1\,0\,32) magnetic reflection in the $\pi$-$\pi$ channel and the (0\,0\,31) reflection in both $\pi$-$\sigma$ and $\pi$-$\pi$ channels can be expressed as,
\begin{eqnarray}
  I_{(1\,0\,4n)}^{\pi-\pi} &=& \left(8\cos\phi\sin\Psi_H\sin2\theta\cos\epsilon\right)^{2} \nonumber \\ 
	I_{(0\,0\,2n+1)}^{\pi-\pi} &=& \left(\sqrt{32}\cos\phi\sin\Psi_H\sin2\theta\cos\epsilon\right)^{2} \nonumber \\ 
	I_{(0\,0\,2n+1)}^{\pi-\sigma} &=& \left(\sqrt{32}\sin\phi\sin\Psi_H\cos\theta\right)^{2} \nonumber 
	\; ,
	\label{eq:intensity}
\end{eqnarray}
where $\theta$ is the scattering angle, $\phi$ the canting angle, and $\Psi_H$ the azimuthal angle of the film with respect to the incident beam (cf. Fig.~\ref{fig:fig3}). $\Psi_H=0$ is defined as the $a$-axis and therefore the magnetic moments of the magnetic $A$-sublattice are in the scattering plane. $\epsilon$ denotes the angle between the scattering vector $\textbf{Q}$ and the $c$-axis (cf.~Fig.~\ref{fig:fig3}). 

\begin{table}
\begin{ruledtabular}
\caption{Parameters used to calculate the expected intensity ratio for a tilting angle $\phi_\mathrm{Bulk}$=12.2$^{\circ}$ of bulk SIO.\cite{Boseggia:2013}}
\label{tab:table2}
\begin{tabular}{|c c c c c c c c|} 
Refl. & Pol.  & $\phi$ & $2\theta$ & $\theta$ & $\epsilon$ &  $\Psi_H$ & Int.\,(a.u.) \\ 
 \hline
 (1\,0\,32) & $\pi$-$\pi$ & $12.2^\circ$ & $87.98^\circ$ & $43.89^\circ$ & $1.79^\circ$ & $90^\circ$ & 60.987 \\ 
 (0\,0\,31) & $\pi$-$\sigma$ & $12.2^\circ$ & $83.48^\circ$ & $41.65^\circ$ & $0^\circ$ & $90^\circ$ & 0.795 \\ 
 (0\,0\,31) & $\pi$-$\pi$ & $12.2^\circ$ & $83.48^\circ$ & $41.65^\circ$ & $0^\circ$ & $0^\circ$ & 1.487 \\ 
\end{tabular}
\end{ruledtabular}
\end{table}
For bulk SIO a tilting angle of $\phi_\mathrm{Bulk}$=12.2$^{\circ}$ has been determined by XRS measurements.\cite{Boseggia:2013} Using the values in Table~\ref{tab:table2}, we expect an intensity ratio of $I_{(1\,0\,32)}^{\pi-\pi}/I_{(0\,0\,31)}^{\pi-\sigma}= 76$ and $I_{(1\,0\,32)}^{\pi-\pi}/I_{(0\,0\,31)}^{\pi-\pi}=41$ for the $\pi$-$\sigma$ and $\pi$-$\pi$ channels, respectively. With an intensity of 220\,cps measured for the (1\,0\,32) reflection (cf. Fig.~\ref{fig:fig11}(a)), we therefore would expect an intensity of around 2.89\,cps (5.36\,cps) above the background for the (0\,0\,31) reflection in the $\pi$-$\sigma$ ($\pi$-$\pi$) channel (blue lines in Fig.~\ref{fig:fig11}(b),(c)). From the measured intensities of our SIO thin film, which is at the limit of the sensitivity of our measurements, we find an intensity ratio of $I(1\,0\,32)/I(0\,0\,31) = 575$. This would correspond to a magnetic canting angle of only $\phi_\mathrm{film}\sim 4.5^{\circ}$. This is surprising, since the finite epitaxial strain in our SIO thin films should only marginally affect the rotation angle $\rho$ of the oxygen octahedra (cf.~table~\ref{tab:table1}). Therefore, only a small change of the canting angle $\phi$ with respect to the unstrained SIO bulk crystal is expected, if we assume a bulk-like magnetoelastic coupling.

However, since we observed a dramatic reduced magnetic correlation length $\xi_{c}$ of the magnetic $A$-sublattice (cf.~Fig.~\ref{fig:fig9}) as well as a high disorder with respect to the rotation of the oxygen octahedra for the 108\,nm SIO thin film (cf.~Fig.~\ref{fig:fig4}), we might also expect that the magnetic net moments $m_\mathrm{net}$ are not well correlated along the $c$-axis of SIO, giving rise to weak and diffuse magnetic scattering of the $(0\,0\,31)$-reflection. 

Another possible explanation for the weak scattering of reflections related to the magnetic $B$-sublattice might be a different magnetic stacking structure compared to that of the bulk SIO compound. However, from the current data we are unable to draw any unambiguous conclusion on this possibility.

In summary, the weak magnetic scattering intensity of the $(0\,0\,31)$ reflection, which is related to magnetic $B$-sublattice of our SIO thin film, reveals a slightly different structure compared to the bulk compound. Since we found a net magnetization in our SIO thin films under applied magnetic field, which is comparable to bulk SIO, we deduce that there must be a finite canting of the magnetic moments in our SIO thin films similar to that of the bulk compound. However, in common with our findings for the oxygen rotation reflections with weak intensities, the canting of the magnetic moments and oxygen rotations might not be highly coherent along the $c$-axis throughout the thickness of the SIO film.

\section{Conclusions}
\label{sec:conclusion}

We have performed a vigorous structural and magnetic investigation of epitaxial thin films of the prototype $J_\mathrm{eff}$ = $\frac{1}{2}$ ground state compound Sr$_2$IrO$_4$ (SIO) by X-ray scattering. To investigate strain and size effects, we examined tensile and compressive strained SIO thin films with thicknesses of 27\,nm-108\,nm fabricated on STO and NGO substrates, respectively. We find that our SIO thin films are epitaxially strained up to a thickness of 108\,nm without clear sign of strain relaxation. These thin films show a strong dependence of the out-of-plane magnetic correlation length on the SIO thin film thickness with a critical thickness of 62.5\,nm separating regions with short and long-range magnetic order along the $c$-axis of SIO. Interestingly, the magnetic correlation length seems to be independent of the strain state of the SIO thin film, since the compressive strained SIO thin film on NGO has a similar correlation length as the tensile-strained SIO thin film on STO with almost the same thickness. Additionally, only very weak intensities of reflections related to the canting of the Ir$^{4+}$-magnetic moments in out-of-plane direction could be observed. This might reflect the disordered oxygen-lattice structure found in the 108\,nm SIO thin film rather than a small canting angle of the Ir$^{4+}$-moments. 

Our work thus brings to light the possibility of fine tuning the magnetic correlation length by using the film thickness as an external parameter to select the desired out-of-plane magnetic correlation length, demonstrating that iridate thin films can be considered as a vast toolkit for the systematic investigation of 5$d$-materials with strong spin-orbit coupling. The tuning of the magnetic correlation length together with electron doping could lead to new routes to the eventual stabilization of superconductivity in the SIO compound.

\begin{acknowledgments}
This work was supported by the German Research Foundation via Germany's  Excellence Strategy (EXC-2111-390814868) and by the European Synchrotron Radiation facility (ESRF) via HC1821 on XMaS and 02-02-855 on D2AM as well as the Diamond Light Source via MT12770. XMaS is a UK National Research Facility funded by EPSRC and managed by the Universities of Liverpool and Warwick. The authors further thank J.~Fischer for support in fabricating SIO thin films as well as T.~Brenninger, A.~Habel, and K.~Helm-Knapp for technical support. DM thanks Tim Ziman for fruitful discussions. 
\end{acknowledgments} 


\begin{thebibliography}{80}%
\makeatletter
\providecommand \@ifxundefined [1]{%
 \@ifx{#1\undefined}
}%
\providecommand \@ifnum [1]{%
 \ifnum #1\expandafter \@firstoftwo
 \else \expandafter \@secondoftwo
 \fi
}%
\providecommand \@ifx [1]{%
 \ifx #1\expandafter \@firstoftwo
 \else \expandafter \@secondoftwo
 \fi
}%
\providecommand \natexlab [1]{#1}%
\providecommand \enquote  [1]{``#1''}%
\providecommand \bibnamefont  [1]{#1}%
\providecommand \bibfnamefont [1]{#1}%
\providecommand \citenamefont [1]{#1}%
\providecommand \href@noop [0]{\@secondoftwo}%
\providecommand \href [0]{\begingroup \@sanitize@url \@href}%
\providecommand \@href[1]{\@@startlink{#1}\@@href}%
\providecommand \@@href[1]{\endgroup#1\@@endlink}%
\providecommand \@sanitize@url [0]{\catcode `\\12\catcode `\$12\catcode
  `\&12\catcode `\#12\catcode `\^12\catcode `\_12\catcode `\%12\relax}%
\providecommand \@@startlink[1]{}%
\providecommand \@@endlink[0]{}%
\providecommand \url  [0]{\begingroup\@sanitize@url \@url }%
\providecommand \@url [1]{\endgroup\@href {#1}{\urlprefix }}%
\providecommand \urlprefix  [0]{URL }%
\providecommand \Eprint [0]{\href }%
\providecommand \doibase [0]{http://dx.doi.org/}%
\providecommand \selectlanguage [0]{\@gobble}%
\providecommand \bibinfo  [0]{\@secondoftwo}%
\providecommand \bibfield  [0]{\@secondoftwo}%
\providecommand \translation [1]{[#1]}%
\providecommand \BibitemOpen [0]{}%
\providecommand \bibitemStop [0]{}%
\providecommand \bibitemNoStop [0]{.\EOS\space}%
\providecommand \EOS [0]{\spacefactor3000\relax}%
\providecommand \BibitemShut  [1]{\csname bibitem#1\endcsname}%
\let\auto@bib@innerbib\@empty
\bibitem [{\citenamefont {Rao}(1989)}]{Rao:1989}%
  \BibitemOpen
  \bibfield  {author} {\bibinfo {author} {\bibfnamefont {C.~N.~R.}\
  \bibnamefont {Rao}},\ }\href@noop {} {\bibfield  {journal} {\bibinfo
  {journal} {Annu. Rev. Phys. Chem.}\ }\textbf {\bibinfo {volume} {40}},\
  \bibinfo {pages} {291} (\bibinfo {year} {1989})}\BibitemShut {NoStop}%
\bibitem [{\citenamefont {Tokura}\ and\ \citenamefont
  {Nagaosa}(2000)}]{Tokura:2000}%
  \BibitemOpen
  \bibfield  {author} {\bibinfo {author} {\bibfnamefont {Y.}~\bibnamefont
  {Tokura}}\ and\ \bibinfo {author} {\bibfnamefont {N.}~\bibnamefont
  {Nagaosa}},\ }\href {\doibase 10.1126/science.288.5465.462} {\bibfield
  {journal} {\bibinfo  {journal} {Science}\ }\textbf {\bibinfo {volume}
  {288}},\ \bibinfo {pages} {462} (\bibinfo {year} {2000})}\BibitemShut
  {NoStop}%
\bibitem [{\citenamefont {Wang}\ \emph {et~al.}(2009)\citenamefont {Wang},
  \citenamefont {Liu},\ and\ \citenamefont {Ren}}]{Wang:2009}%
  \BibitemOpen
  \bibfield  {author} {\bibinfo {author} {\bibfnamefont {K.}~\bibnamefont
  {Wang}}, \bibinfo {author} {\bibfnamefont {J.-M.}\ \bibnamefont {Liu}}, \
  and\ \bibinfo {author} {\bibfnamefont {Z.}~\bibnamefont {Ren}},\ }\href
  {\doibase 10.1080/00018730902920554} {\bibfield  {journal} {\bibinfo
  {journal} {Adv. Phy.}\ }\textbf {\bibinfo {volume} {58}},\ \bibinfo {pages}
  {321} (\bibinfo {year} {2009})}\BibitemShut {NoStop}%
\bibitem [{\citenamefont {Imada}\ \emph {et~al.}(1998)\citenamefont {Imada},
  \citenamefont {Fujimori},\ and\ \citenamefont {Tokura}}]{Imada:1998}%
  \BibitemOpen
  \bibfield  {author} {\bibinfo {author} {\bibfnamefont {M.}~\bibnamefont
  {Imada}}, \bibinfo {author} {\bibfnamefont {A.}~\bibnamefont {Fujimori}}, \
  and\ \bibinfo {author} {\bibfnamefont {Y.}~\bibnamefont {Tokura}},\ }\href
  {\doibase 10.1103/RevModPhys.70.1039} {\bibfield  {journal} {\bibinfo
  {journal} {Rev. Mod. Phys.}\ }\textbf {\bibinfo {volume} {70}},\ \bibinfo
  {pages} {1039} (\bibinfo {year} {1998})}\BibitemShut {NoStop}%
\bibitem [{\citenamefont {Urushibara}\ \emph {et~al.}(1995)\citenamefont
  {Urushibara}, \citenamefont {Moritomo}, \citenamefont {Arima}, \citenamefont
  {Asamitsu}, \citenamefont {Kido},\ and\ \citenamefont
  {Tokura}}]{Urushibara:1995}%
  \BibitemOpen
  \bibfield  {author} {\bibinfo {author} {\bibfnamefont {A.}~\bibnamefont
  {Urushibara}}, \bibinfo {author} {\bibfnamefont {Y.}~\bibnamefont
  {Moritomo}}, \bibinfo {author} {\bibfnamefont {T.}~\bibnamefont {Arima}},
  \bibinfo {author} {\bibfnamefont {A.}~\bibnamefont {Asamitsu}}, \bibinfo
  {author} {\bibfnamefont {G.}~\bibnamefont {Kido}}, \ and\ \bibinfo {author}
  {\bibfnamefont {Y.}~\bibnamefont {Tokura}},\ }\href {\doibase
  10.1103/PhysRevB.51.14103} {\bibfield  {journal} {\bibinfo  {journal} {Phys.
  Rev. B}\ }\textbf {\bibinfo {volume} {51}},\ \bibinfo {pages} {14103}
  (\bibinfo {year} {1995})}\BibitemShut {NoStop}%
\bibitem [{\citenamefont {Bednorz}\ and\ \citenamefont
  {M\"uller}(1986)}]{Bednorz:1986}%
  \BibitemOpen
  \bibfield  {author} {\bibinfo {author} {\bibfnamefont {J.~G.}\ \bibnamefont
  {Bednorz}}\ and\ \bibinfo {author} {\bibfnamefont {K.~A.}\ \bibnamefont
  {M\"uller}},\ }\href {https://doi.org/10.1007/BF01303701} {\bibfield
  {journal} {\bibinfo  {journal} {Z. Physik B - Condensed Matter}\ }\textbf
  {\bibinfo {volume} {64}},\ \bibinfo {pages} {189} (\bibinfo {year}
  {1986})}\BibitemShut {NoStop}%
\bibitem [{\citenamefont {Pesin}\ and\ \citenamefont
  {Balents}(2010)}]{Pesin:2010}%
  \BibitemOpen
  \bibfield  {author} {\bibinfo {author} {\bibfnamefont {D.}~\bibnamefont
  {Pesin}}\ and\ \bibinfo {author} {\bibfnamefont {L.}~\bibnamefont
  {Balents}},\ }\href {https://doi.org/10.1038/nphys1606} {\bibfield  {journal}
  {\bibinfo  {journal} {Nat. Phys.}\ }\textbf {\bibinfo {volume} {6}},\
  \bibinfo {pages} {376} (\bibinfo {year} {2010})}\BibitemShut {NoStop}%
\bibitem [{\citenamefont {Jackeli}\ and\ \citenamefont
  {Khaliullin}(2009)}]{Jackeli:2009}%
  \BibitemOpen
  \bibfield  {author} {\bibinfo {author} {\bibfnamefont {G.}~\bibnamefont
  {Jackeli}}\ and\ \bibinfo {author} {\bibfnamefont {G.}~\bibnamefont
  {Khaliullin}},\ }\href {\doibase 10.1103/PhysRevLett.102.017205} {\bibfield
  {journal} {\bibinfo  {journal} {Phys. Rev. Lett.}\ }\textbf {\bibinfo
  {volume} {102}},\ \bibinfo {pages} {017205} (\bibinfo {year}
  {2009})}\BibitemShut {NoStop}%
\bibitem [{\citenamefont {Wan}\ \emph {et~al.}(2011)\citenamefont {Wan},
  \citenamefont {Turner}, \citenamefont {Vishwanath},\ and\ \citenamefont
  {Savrasov}}]{Wan:2011}%
  \BibitemOpen
  \bibfield  {author} {\bibinfo {author} {\bibfnamefont {X.}~\bibnamefont
  {Wan}}, \bibinfo {author} {\bibfnamefont {A.~M.}\ \bibnamefont {Turner}},
  \bibinfo {author} {\bibfnamefont {A.}~\bibnamefont {Vishwanath}}, \ and\
  \bibinfo {author} {\bibfnamefont {S.~Y.}\ \bibnamefont {Savrasov}},\ }\href
  {\doibase 10.1103/PhysRevB.83.205101} {\bibfield  {journal} {\bibinfo
  {journal} {Phys. Rev. B}\ }\textbf {\bibinfo {volume} {83}},\ \bibinfo
  {pages} {205101} (\bibinfo {year} {2011})}\BibitemShut {NoStop}%
\bibitem [{\citenamefont {Witczak-Krempa}\ \emph {et~al.}(2014)\citenamefont
  {Witczak-Krempa}, \citenamefont {Chen}, \citenamefont {Kim},\ and\
  \citenamefont {Balents}}]{Krempa:2014}%
  \BibitemOpen
  \bibfield  {author} {\bibinfo {author} {\bibfnamefont {W.}~\bibnamefont
  {Witczak-Krempa}}, \bibinfo {author} {\bibfnamefont {G.}~\bibnamefont
  {Chen}}, \bibinfo {author} {\bibfnamefont {Y.~B.}\ \bibnamefont {Kim}}, \
  and\ \bibinfo {author} {\bibfnamefont {L.}~\bibnamefont {Balents}},\ }\href
  {\doibase 10.1146/annurev-conmatphys-020911-125138} {\bibfield  {journal}
  {\bibinfo  {journal} {Annu. Rev. Condens. Matter Phys.}\ }\textbf {\bibinfo
  {volume} {5}},\ \bibinfo {pages} {57} (\bibinfo {year} {2014})}\BibitemShut
  {NoStop}%
\bibitem [{\citenamefont {Kim}\ \emph {et~al.}(2008)\citenamefont {Kim},
  \citenamefont {Jin}, \citenamefont {Moon}, \citenamefont {Kim}, \citenamefont
  {Park}, \citenamefont {Leem}, \citenamefont {Yu}, \citenamefont {Noh},
  \citenamefont {Kim}, \citenamefont {Oh}, \citenamefont {Park}, \citenamefont
  {Durairaj}, \citenamefont {Cao},\ and\ \citenamefont {Rotenberg}}]{Kim:2008}%
  \BibitemOpen
  \bibfield  {author} {\bibinfo {author} {\bibfnamefont {B.~J.}\ \bibnamefont
  {Kim}}, \bibinfo {author} {\bibfnamefont {H.}~\bibnamefont {Jin}}, \bibinfo
  {author} {\bibfnamefont {S.~J.}\ \bibnamefont {Moon}}, \bibinfo {author}
  {\bibfnamefont {J.-Y.}\ \bibnamefont {Kim}}, \bibinfo {author} {\bibfnamefont
  {B.-G.}\ \bibnamefont {Park}}, \bibinfo {author} {\bibfnamefont {C.~S.}\
  \bibnamefont {Leem}}, \bibinfo {author} {\bibfnamefont {J.}~\bibnamefont
  {Yu}}, \bibinfo {author} {\bibfnamefont {T.~W.}\ \bibnamefont {Noh}},
  \bibinfo {author} {\bibfnamefont {C.}~\bibnamefont {Kim}}, \bibinfo {author}
  {\bibfnamefont {S.-J.}\ \bibnamefont {Oh}}, \bibinfo {author} {\bibfnamefont
  {J.-H.}\ \bibnamefont {Park}}, \bibinfo {author} {\bibfnamefont
  {V.}~\bibnamefont {Durairaj}}, \bibinfo {author} {\bibfnamefont
  {G.}~\bibnamefont {Cao}}, \ and\ \bibinfo {author} {\bibfnamefont
  {E.}~\bibnamefont {Rotenberg}},\ }\href {\doibase
  10.1103/PhysRevLett.101.076402} {\bibfield  {journal} {\bibinfo  {journal}
  {Phys. Rev. Lett.}\ }\textbf {\bibinfo {volume} {101}},\ \bibinfo {pages}
  {076402} (\bibinfo {year} {2008})}\BibitemShut {NoStop}%
\bibitem [{\citenamefont {Moon}\ \emph {et~al.}(2008)\citenamefont {Moon},
  \citenamefont {Jin}, \citenamefont {Kim}, \citenamefont {Choi}, \citenamefont
  {Lee}, \citenamefont {Yu}, \citenamefont {Cao}, \citenamefont {Sumi},
  \citenamefont {Funakubo}, \citenamefont {Bernhard},\ and\ \citenamefont
  {Noh}}]{Moon:2008}%
  \BibitemOpen
  \bibfield  {author} {\bibinfo {author} {\bibfnamefont {S.~J.}\ \bibnamefont
  {Moon}}, \bibinfo {author} {\bibfnamefont {H.}~\bibnamefont {Jin}}, \bibinfo
  {author} {\bibfnamefont {K.~W.}\ \bibnamefont {Kim}}, \bibinfo {author}
  {\bibfnamefont {W.~S.}\ \bibnamefont {Choi}}, \bibinfo {author}
  {\bibfnamefont {Y.~S.}\ \bibnamefont {Lee}}, \bibinfo {author} {\bibfnamefont
  {J.}~\bibnamefont {Yu}}, \bibinfo {author} {\bibfnamefont {G.}~\bibnamefont
  {Cao}}, \bibinfo {author} {\bibfnamefont {A.}~\bibnamefont {Sumi}}, \bibinfo
  {author} {\bibfnamefont {H.}~\bibnamefont {Funakubo}}, \bibinfo {author}
  {\bibfnamefont {C.}~\bibnamefont {Bernhard}}, \ and\ \bibinfo {author}
  {\bibfnamefont {T.~W.}\ \bibnamefont {Noh}},\ }\href {\doibase
  10.1103/PhysRevLett.101.226402} {\bibfield  {journal} {\bibinfo  {journal}
  {Phys. Rev. Lett.}\ }\textbf {\bibinfo {volume} {101}},\ \bibinfo {pages}
  {226402} (\bibinfo {year} {2008})}\BibitemShut {NoStop}%
\bibitem [{\citenamefont {Kim}\ \emph {et~al.}(2009)\citenamefont {Kim},
  \citenamefont {Ohsumi}, \citenamefont {Komesu}, \citenamefont {Sakai},
  \citenamefont {Morita}, \citenamefont {Takagi},\ and\ \citenamefont
  {Arima}}]{Kim:2009}%
  \BibitemOpen
  \bibfield  {author} {\bibinfo {author} {\bibfnamefont {B.~J.}\ \bibnamefont
  {Kim}}, \bibinfo {author} {\bibfnamefont {H.}~\bibnamefont {Ohsumi}},
  \bibinfo {author} {\bibfnamefont {T.}~\bibnamefont {Komesu}}, \bibinfo
  {author} {\bibfnamefont {S.}~\bibnamefont {Sakai}}, \bibinfo {author}
  {\bibfnamefont {T.}~\bibnamefont {Morita}}, \bibinfo {author} {\bibfnamefont
  {H.}~\bibnamefont {Takagi}}, \ and\ \bibinfo {author} {\bibfnamefont
  {T.}~\bibnamefont {Arima}},\ }\href {\doibase 10.1126/science.1167106}
  {\bibfield  {journal} {\bibinfo  {journal} {Science}\ }\textbf {\bibinfo
  {volume} {323}},\ \bibinfo {pages} {1329} (\bibinfo {year}
  {2009})}\BibitemShut {NoStop}%
\bibitem [{\citenamefont {Lu}\ and\ \citenamefont {Liu}(2020)}]{Lu:2020}%
  \BibitemOpen
  \bibfield  {author} {\bibinfo {author} {\bibfnamefont {C.}~\bibnamefont
  {Lu}}\ and\ \bibinfo {author} {\bibfnamefont {J.-M.}\ \bibnamefont {Liu}},\
  }\href {\doibase 10.1002/adma.201904508} {\bibfield  {journal} {\bibinfo
  {journal} {Adv. Mater.}\ }\textbf {\bibinfo {volume} {32}},\ \bibinfo {pages}
  {1904508} (\bibinfo {year} {2020})}\BibitemShut {NoStop}%
\bibitem [{\citenamefont {Crawford}\ \emph {et~al.}(1994)\citenamefont
  {Crawford}, \citenamefont {Subramanian}, \citenamefont {Harlow},
  \citenamefont {Fernandez-Baca}, \citenamefont {Wang},\ and\ \citenamefont
  {Johnston}}]{Crawford:1994}%
  \BibitemOpen
  \bibfield  {author} {\bibinfo {author} {\bibfnamefont {M.~K.}\ \bibnamefont
  {Crawford}}, \bibinfo {author} {\bibfnamefont {M.~A.}\ \bibnamefont
  {Subramanian}}, \bibinfo {author} {\bibfnamefont {R.~L.}\ \bibnamefont
  {Harlow}}, \bibinfo {author} {\bibfnamefont {J.~A.}\ \bibnamefont
  {Fernandez-Baca}}, \bibinfo {author} {\bibfnamefont {Z.~R.}\ \bibnamefont
  {Wang}}, \ and\ \bibinfo {author} {\bibfnamefont {D.~C.}\ \bibnamefont
  {Johnston}},\ }\href {\doibase 10.1103/PhysRevB.49.9198} {\bibfield
  {journal} {\bibinfo  {journal} {Phys. Rev. B}\ }\textbf {\bibinfo {volume}
  {49}},\ \bibinfo {pages} {9198} (\bibinfo {year} {1994})}\BibitemShut
  {NoStop}%
\bibitem [{\citenamefont {Kim}\ \emph {et~al.}(2014)\citenamefont {Kim},
  \citenamefont {Krupin}, \citenamefont {Denlinger}, \citenamefont {Bostwick},
  \citenamefont {Rotenberg}, \citenamefont {Zhao}, \citenamefont {Mitchell},
  \citenamefont {Allen},\ and\ \citenamefont {Kim}}]{Kim:2014}%
  \BibitemOpen
  \bibfield  {author} {\bibinfo {author} {\bibfnamefont {Y.~K.}\ \bibnamefont
  {Kim}}, \bibinfo {author} {\bibfnamefont {O.}~\bibnamefont {Krupin}},
  \bibinfo {author} {\bibfnamefont {J.~D.}\ \bibnamefont {Denlinger}}, \bibinfo
  {author} {\bibfnamefont {A.}~\bibnamefont {Bostwick}}, \bibinfo {author}
  {\bibfnamefont {E.}~\bibnamefont {Rotenberg}}, \bibinfo {author}
  {\bibfnamefont {Q.}~\bibnamefont {Zhao}}, \bibinfo {author} {\bibfnamefont
  {J.~F.}\ \bibnamefont {Mitchell}}, \bibinfo {author} {\bibfnamefont {J.~W.}\
  \bibnamefont {Allen}}, \ and\ \bibinfo {author} {\bibfnamefont {B.~J.}\
  \bibnamefont {Kim}},\ }\href {\doibase 10.1126/science.1251151} {\bibfield
  {journal} {\bibinfo  {journal} {Science}\ }\textbf {\bibinfo {volume}
  {345}},\ \bibinfo {pages} {187} (\bibinfo {year} {2014})}\BibitemShut
  {NoStop}%
\bibitem [{\citenamefont {Yan}\ \emph {et~al.}(2015)\citenamefont {Yan},
  \citenamefont {Ren}, \citenamefont {Xu}, \citenamefont {Xie}, \citenamefont
  {Tao}, \citenamefont {Choi}, \citenamefont {Lee}, \citenamefont {Choi},
  \citenamefont {Zhang},\ and\ \citenamefont {Feng}}]{Yan:2015}%
  \BibitemOpen
  \bibfield  {author} {\bibinfo {author} {\bibfnamefont {Y.~J.}\ \bibnamefont
  {Yan}}, \bibinfo {author} {\bibfnamefont {M.~Q.}\ \bibnamefont {Ren}},
  \bibinfo {author} {\bibfnamefont {H.~C.}\ \bibnamefont {Xu}}, \bibinfo
  {author} {\bibfnamefont {B.~P.}\ \bibnamefont {Xie}}, \bibinfo {author}
  {\bibfnamefont {R.}~\bibnamefont {Tao}}, \bibinfo {author} {\bibfnamefont
  {H.~Y.}\ \bibnamefont {Choi}}, \bibinfo {author} {\bibfnamefont
  {N.}~\bibnamefont {Lee}}, \bibinfo {author} {\bibfnamefont {Y.~J.}\
  \bibnamefont {Choi}}, \bibinfo {author} {\bibfnamefont {T.}~\bibnamefont
  {Zhang}}, \ and\ \bibinfo {author} {\bibfnamefont {D.~L.}\ \bibnamefont
  {Feng}},\ }\href {\doibase 10.1103/PhysRevX.5.041018} {\bibfield  {journal}
  {\bibinfo  {journal} {Phys. Rev. X}\ }\textbf {\bibinfo {volume} {5}},\
  \bibinfo {pages} {041018} (\bibinfo {year} {2015})}\BibitemShut {NoStop}%
\bibitem [{\citenamefont {Battisti}\ \emph {et~al.}(2016)\citenamefont
  {Battisti}, \citenamefont {Bastiaans}, \citenamefont {Fedoseev},
  \citenamefont {de~la Torre}, \citenamefont {Iliopoulos}, \citenamefont
  {Tamai}, \citenamefont {Hunter}, \citenamefont {Perry}, \citenamefont
  {Zaanen}, \citenamefont {Baumberger},\ and\ \citenamefont
  {Allan}}]{Battisti:2016}%
  \BibitemOpen
  \bibfield  {author} {\bibinfo {author} {\bibfnamefont {I.}~\bibnamefont
  {Battisti}}, \bibinfo {author} {\bibfnamefont {K.~M.}\ \bibnamefont
  {Bastiaans}}, \bibinfo {author} {\bibfnamefont {V.}~\bibnamefont {Fedoseev}},
  \bibinfo {author} {\bibfnamefont {A.}~\bibnamefont {de~la Torre}}, \bibinfo
  {author} {\bibfnamefont {N.}~\bibnamefont {Iliopoulos}}, \bibinfo {author}
  {\bibfnamefont {A.}~\bibnamefont {Tamai}}, \bibinfo {author} {\bibfnamefont
  {E.~C.}\ \bibnamefont {Hunter}}, \bibinfo {author} {\bibfnamefont
  {R.}~\bibnamefont {Perry}}, \bibinfo {author} {\bibfnamefont
  {J.}~\bibnamefont {Zaanen}}, \bibinfo {author} {\bibfnamefont
  {F.}~\bibnamefont {Baumberger}}, \ and\ \bibinfo {author} {\bibfnamefont
  {M.~P.}\ \bibnamefont {Allan}},\ }\href {https://doi.org/10.1038/nphys3894}
  {\bibfield  {journal} {\bibinfo  {journal} {Nat. Phys.}\ }\textbf {\bibinfo
  {volume} {13}},\ \bibinfo {pages} {21} (\bibinfo {year} {2016})}\BibitemShut
  {NoStop}%
\bibitem [{\citenamefont {Kim}\ \emph {et~al.}(2016{\natexlab{a}})\citenamefont
  {Kim}, \citenamefont {Sung}, \citenamefont {Denlinger},\ and\ \citenamefont
  {Kim}}]{Kim:2016}%
  \BibitemOpen
  \bibfield  {author} {\bibinfo {author} {\bibfnamefont {Y.~K.}\ \bibnamefont
  {Kim}}, \bibinfo {author} {\bibfnamefont {N.~H.}\ \bibnamefont {Sung}},
  \bibinfo {author} {\bibfnamefont {J.~D.}\ \bibnamefont {Denlinger}}, \ and\
  \bibinfo {author} {\bibfnamefont {B.~J.}\ \bibnamefont {Kim}},\ }\href
  {https://doi.org/10.1038/nphys3503} {\bibfield  {journal} {\bibinfo
  {journal} {Nat. Phys.}\ }\textbf {\bibinfo {volume} {12}},\ \bibinfo {pages}
  {37} (\bibinfo {year} {2016}{\natexlab{a}})}\BibitemShut {NoStop}%
\bibitem [{\citenamefont {Zhao}\ \emph {et~al.}(2015)\citenamefont {Zhao},
  \citenamefont {Torchinsky}, \citenamefont {Chu}, \citenamefont {Ivanov},
  \citenamefont {Lifshitz}, \citenamefont {Flint}, \citenamefont {Qi},
  \citenamefont {Cao},\ and\ \citenamefont {Hsieh}}]{Zhao:2015}%
  \BibitemOpen
  \bibfield  {author} {\bibinfo {author} {\bibfnamefont {L.}~\bibnamefont
  {Zhao}}, \bibinfo {author} {\bibfnamefont {D.~H.}\ \bibnamefont
  {Torchinsky}}, \bibinfo {author} {\bibfnamefont {H.}~\bibnamefont {Chu}},
  \bibinfo {author} {\bibfnamefont {V.}~\bibnamefont {Ivanov}}, \bibinfo
  {author} {\bibfnamefont {R.}~\bibnamefont {Lifshitz}}, \bibinfo {author}
  {\bibfnamefont {R.}~\bibnamefont {Flint}}, \bibinfo {author} {\bibfnamefont
  {T.}~\bibnamefont {Qi}}, \bibinfo {author} {\bibfnamefont {G.}~\bibnamefont
  {Cao}}, \ and\ \bibinfo {author} {\bibfnamefont {D.}~\bibnamefont {Hsieh}},\
  }\href {https://doi.org/10.1038/nphys3517} {\bibfield  {journal} {\bibinfo
  {journal} {Nat. Phys.}\ }\textbf {\bibinfo {volume} {12}},\ \bibinfo {pages}
  {32} (\bibinfo {year} {2015})}\BibitemShut {NoStop}%
\bibitem [{\citenamefont {Jeong}\ \emph {et~al.}(2017)\citenamefont {Jeong},
  \citenamefont {Sidis}, \citenamefont {Louat}, \citenamefont {Brouet},\ and\
  \citenamefont {Bourges}}]{Jeong:2017}%
  \BibitemOpen
  \bibfield  {author} {\bibinfo {author} {\bibfnamefont {J.}~\bibnamefont
  {Jeong}}, \bibinfo {author} {\bibfnamefont {Y.}~\bibnamefont {Sidis}},
  \bibinfo {author} {\bibfnamefont {A.}~\bibnamefont {Louat}}, \bibinfo
  {author} {\bibfnamefont {V.}~\bibnamefont {Brouet}}, \ and\ \bibinfo {author}
  {\bibfnamefont {P.}~\bibnamefont {Bourges}},\ }\href
  {https://doi.org/10.1038/ncomms15119} {\bibfield  {journal} {\bibinfo
  {journal} {Nature Comm.}\ }\textbf {\bibinfo {volume} {8}},\ \bibinfo {pages}
  {15119} (\bibinfo {year} {2017})}\BibitemShut {NoStop}%
\bibitem [{\citenamefont {Qi}\ \emph {et~al.}(2011)\citenamefont {Qi},
  \citenamefont {Korneta}, \citenamefont {Chikara}, \citenamefont {Ge},
  \citenamefont {Parkin}, \citenamefont {De~Long}, \citenamefont
  {Schlottmann},\ and\ \citenamefont {Cao}}]{Qi:2011}%
  \BibitemOpen
  \bibfield  {author} {\bibinfo {author} {\bibfnamefont {T.~F.}\ \bibnamefont
  {Qi}}, \bibinfo {author} {\bibfnamefont {O.~B.}\ \bibnamefont {Korneta}},
  \bibinfo {author} {\bibfnamefont {S.}~\bibnamefont {Chikara}}, \bibinfo
  {author} {\bibfnamefont {M.}~\bibnamefont {Ge}}, \bibinfo {author}
  {\bibfnamefont {S.}~\bibnamefont {Parkin}}, \bibinfo {author} {\bibfnamefont
  {L.~E.}\ \bibnamefont {De~Long}}, \bibinfo {author} {\bibfnamefont
  {P.}~\bibnamefont {Schlottmann}}, \ and\ \bibinfo {author} {\bibfnamefont
  {G.}~\bibnamefont {Cao}},\ }\href {\doibase 10.1063/1.3545803} {\bibfield
  {journal} {\bibinfo  {journal} {J. Appl. Phys.}\ }\textbf {\bibinfo {volume}
  {109}},\ \bibinfo {pages} {07D906} (\bibinfo {year} {2011})}\BibitemShut
  {NoStop}%
\bibitem [{\citenamefont {Watanabe}\ \emph {et~al.}(2013)\citenamefont
  {Watanabe}, \citenamefont {Shirakawa},\ and\ \citenamefont
  {Yunoki}}]{Watanabe:2013}%
  \BibitemOpen
  \bibfield  {author} {\bibinfo {author} {\bibfnamefont {H.}~\bibnamefont
  {Watanabe}}, \bibinfo {author} {\bibfnamefont {T.}~\bibnamefont {Shirakawa}},
  \ and\ \bibinfo {author} {\bibfnamefont {S.}~\bibnamefont {Yunoki}},\ }\href
  {\doibase 10.1103/PhysRevLett.110.027002} {\bibfield  {journal} {\bibinfo
  {journal} {Phys. Rev. Lett.}\ }\textbf {\bibinfo {volume} {110}},\ \bibinfo
  {pages} {027002} (\bibinfo {year} {2013})}\BibitemShut {NoStop}%
\bibitem [{\citenamefont {Meng}\ \emph {et~al.}(2014)\citenamefont {Meng},
  \citenamefont {Kim},\ and\ \citenamefont {Kee}}]{Meng:2014}%
  \BibitemOpen
  \bibfield  {author} {\bibinfo {author} {\bibfnamefont {Z.~Y.}\ \bibnamefont
  {Meng}}, \bibinfo {author} {\bibfnamefont {Y.~B.}\ \bibnamefont {Kim}}, \
  and\ \bibinfo {author} {\bibfnamefont {H.-Y.}\ \bibnamefont {Kee}},\ }\href
  {\doibase 10.1103/PhysRevLett.113.177003} {\bibfield  {journal} {\bibinfo
  {journal} {Phys. Rev. Lett.}\ }\textbf {\bibinfo {volume} {113}},\ \bibinfo
  {pages} {177003} (\bibinfo {year} {2014})}\BibitemShut {NoStop}%
\bibitem [{\citenamefont {Dhital}\ \emph {et~al.}(2013)\citenamefont {Dhital},
  \citenamefont {Hogan}, \citenamefont {Yamani}, \citenamefont {de~la Cruz},
  \citenamefont {Chen}, \citenamefont {Khadka}, \citenamefont {Ren},\ and\
  \citenamefont {Wilson}}]{Dhital:2013}%
  \BibitemOpen
  \bibfield  {author} {\bibinfo {author} {\bibfnamefont {C.}~\bibnamefont
  {Dhital}}, \bibinfo {author} {\bibfnamefont {T.}~\bibnamefont {Hogan}},
  \bibinfo {author} {\bibfnamefont {Z.}~\bibnamefont {Yamani}}, \bibinfo
  {author} {\bibfnamefont {C.}~\bibnamefont {de~la Cruz}}, \bibinfo {author}
  {\bibfnamefont {X.}~\bibnamefont {Chen}}, \bibinfo {author} {\bibfnamefont
  {S.}~\bibnamefont {Khadka}}, \bibinfo {author} {\bibfnamefont
  {Z.}~\bibnamefont {Ren}}, \ and\ \bibinfo {author} {\bibfnamefont {S.~D.}\
  \bibnamefont {Wilson}},\ }\href {\doibase 10.1103/PhysRevB.87.144405}
  {\bibfield  {journal} {\bibinfo  {journal} {Phys. Rev. B}\ }\textbf {\bibinfo
  {volume} {87}},\ \bibinfo {pages} {144405} (\bibinfo {year}
  {2013})}\BibitemShut {NoStop}%
\bibitem [{\citenamefont {Ye}\ \emph {et~al.}(2015)\citenamefont {Ye},
  \citenamefont {Wang}, \citenamefont {Hoffmann}, \citenamefont {Wang},
  \citenamefont {Chi}, \citenamefont {Matsuda}, \citenamefont {Chakoumakos},
  \citenamefont {Fernandez-Baca},\ and\ \citenamefont {Cao}}]{Ye:2015}%
  \BibitemOpen
  \bibfield  {author} {\bibinfo {author} {\bibfnamefont {F.}~\bibnamefont
  {Ye}}, \bibinfo {author} {\bibfnamefont {X.}~\bibnamefont {Wang}}, \bibinfo
  {author} {\bibfnamefont {C.}~\bibnamefont {Hoffmann}}, \bibinfo {author}
  {\bibfnamefont {J.}~\bibnamefont {Wang}}, \bibinfo {author} {\bibfnamefont
  {S.}~\bibnamefont {Chi}}, \bibinfo {author} {\bibfnamefont {M.}~\bibnamefont
  {Matsuda}}, \bibinfo {author} {\bibfnamefont {B.~C.}\ \bibnamefont
  {Chakoumakos}}, \bibinfo {author} {\bibfnamefont {J.~A.}\ \bibnamefont
  {Fernandez-Baca}}, \ and\ \bibinfo {author} {\bibfnamefont {G.}~\bibnamefont
  {Cao}},\ }\href {\doibase 10.1103/PhysRevB.92.201112} {\bibfield  {journal}
  {\bibinfo  {journal} {Phys. Rev. B}\ }\textbf {\bibinfo {volume} {92}},\
  \bibinfo {pages} {201112} (\bibinfo {year} {2015})}\BibitemShut {NoStop}%
\bibitem [{\citenamefont {Torchinsky}\ \emph {et~al.}(2015)\citenamefont
  {Torchinsky}, \citenamefont {Chu}, \citenamefont {Zhao}, \citenamefont
  {Perkins}, \citenamefont {Sizyuk}, \citenamefont {Qi}, \citenamefont {Cao},\
  and\ \citenamefont {Hsieh}}]{Torchinsky:2015}%
  \BibitemOpen
  \bibfield  {author} {\bibinfo {author} {\bibfnamefont {D.~H.}\ \bibnamefont
  {Torchinsky}}, \bibinfo {author} {\bibfnamefont {H.}~\bibnamefont {Chu}},
  \bibinfo {author} {\bibfnamefont {L.}~\bibnamefont {Zhao}}, \bibinfo {author}
  {\bibfnamefont {N.~B.}\ \bibnamefont {Perkins}}, \bibinfo {author}
  {\bibfnamefont {Y.}~\bibnamefont {Sizyuk}}, \bibinfo {author} {\bibfnamefont
  {T.}~\bibnamefont {Qi}}, \bibinfo {author} {\bibfnamefont {G.}~\bibnamefont
  {Cao}}, \ and\ \bibinfo {author} {\bibfnamefont {D.}~\bibnamefont {Hsieh}},\
  }\href {\doibase 10.1103/PhysRevLett.114.096404} {\bibfield  {journal}
  {\bibinfo  {journal} {Phys. Rev. Lett.}\ }\textbf {\bibinfo {volume} {114}},\
  \bibinfo {pages} {096404} (\bibinfo {year} {2015})}\BibitemShut {NoStop}%
\bibitem [{\citenamefont {Boseggia}\ \emph
  {et~al.}(2013{\natexlab{a}})\citenamefont {Boseggia}, \citenamefont {Walker},
  \citenamefont {Vale}, \citenamefont {Springell}, \citenamefont {Feng},
  \citenamefont {Perry}, \citenamefont {Sala}, \citenamefont {Rønnow},
  \citenamefont {Collins},\ and\ \citenamefont {McMorrow}}]{Boseggia:2013}%
  \BibitemOpen
  \bibfield  {author} {\bibinfo {author} {\bibfnamefont {S.}~\bibnamefont
  {Boseggia}}, \bibinfo {author} {\bibfnamefont {H.~C.}\ \bibnamefont
  {Walker}}, \bibinfo {author} {\bibfnamefont {J.}~\bibnamefont {Vale}},
  \bibinfo {author} {\bibfnamefont {R.}~\bibnamefont {Springell}}, \bibinfo
  {author} {\bibfnamefont {Z.}~\bibnamefont {Feng}}, \bibinfo {author}
  {\bibfnamefont {R.~S.}\ \bibnamefont {Perry}}, \bibinfo {author}
  {\bibfnamefont {M.~M.}\ \bibnamefont {Sala}}, \bibinfo {author}
  {\bibfnamefont {H.~M.}\ \bibnamefont {Rønnow}}, \bibinfo {author}
  {\bibfnamefont {S.~P.}\ \bibnamefont {Collins}}, \ and\ \bibinfo {author}
  {\bibfnamefont {D.~F.}\ \bibnamefont {McMorrow}},\ }\href
  {http://stacks.iop.org/0953-8984/25/i=42/a=422202} {\bibfield  {journal}
  {\bibinfo  {journal} {J. Phys.: Condens. Matter}\ }\textbf {\bibinfo {volume}
  {25}},\ \bibinfo {pages} {422202} (\bibinfo {year}
  {2013}{\natexlab{a}})}\BibitemShut {NoStop}%
\bibitem [{\citenamefont {Ye}\ \emph {et~al.}(2013)\citenamefont {Ye},
  \citenamefont {Chi}, \citenamefont {Chakoumakos}, \citenamefont
  {Fernandez-Baca}, \citenamefont {Qi},\ and\ \citenamefont {Cao}}]{Ye:2013}%
  \BibitemOpen
  \bibfield  {author} {\bibinfo {author} {\bibfnamefont {F.}~\bibnamefont
  {Ye}}, \bibinfo {author} {\bibfnamefont {S.}~\bibnamefont {Chi}}, \bibinfo
  {author} {\bibfnamefont {B.~C.}\ \bibnamefont {Chakoumakos}}, \bibinfo
  {author} {\bibfnamefont {J.~A.}\ \bibnamefont {Fernandez-Baca}}, \bibinfo
  {author} {\bibfnamefont {T.}~\bibnamefont {Qi}}, \ and\ \bibinfo {author}
  {\bibfnamefont {G.}~\bibnamefont {Cao}},\ }\href {\doibase
  10.1103/PhysRevB.87.140406} {\bibfield  {journal} {\bibinfo  {journal} {Phys.
  Rev. B}\ }\textbf {\bibinfo {volume} {87}},\ \bibinfo {pages} {140406}
  (\bibinfo {year} {2013})}\BibitemShut {NoStop}%
\bibitem [{\citenamefont {Shimura}\ \emph {et~al.}(1995)\citenamefont
  {Shimura}, \citenamefont {Inaguma}, \citenamefont {Nakamura}, \citenamefont
  {Itoh},\ and\ \citenamefont {Morii}}]{Shimura:1995}%
  \BibitemOpen
  \bibfield  {author} {\bibinfo {author} {\bibfnamefont {T.}~\bibnamefont
  {Shimura}}, \bibinfo {author} {\bibfnamefont {Y.}~\bibnamefont {Inaguma}},
  \bibinfo {author} {\bibfnamefont {T.}~\bibnamefont {Nakamura}}, \bibinfo
  {author} {\bibfnamefont {M.}~\bibnamefont {Itoh}}, \ and\ \bibinfo {author}
  {\bibfnamefont {Y.}~\bibnamefont {Morii}},\ }\href {\doibase
  10.1103/PhysRevB.52.9143} {\bibfield  {journal} {\bibinfo  {journal} {Phys.
  Rev. B}\ }\textbf {\bibinfo {volume} {52}},\ \bibinfo {pages} {9143}
  (\bibinfo {year} {1995})}\BibitemShut {NoStop}%
\bibitem [{\citenamefont {Cao}\ \emph {et~al.}(1998)\citenamefont {Cao},
  \citenamefont {Bolivar}, \citenamefont {McCall}, \citenamefont {Crow},\ and\
  \citenamefont {Guertin}}]{Cao:1998}%
  \BibitemOpen
  \bibfield  {author} {\bibinfo {author} {\bibfnamefont {G.}~\bibnamefont
  {Cao}}, \bibinfo {author} {\bibfnamefont {J.}~\bibnamefont {Bolivar}},
  \bibinfo {author} {\bibfnamefont {S.}~\bibnamefont {McCall}}, \bibinfo
  {author} {\bibfnamefont {J.}~\bibnamefont {Crow}}, \ and\ \bibinfo {author}
  {\bibfnamefont {R.}~\bibnamefont {Guertin}},\ }\href@noop {} {\bibfield
  {journal} {\bibinfo  {journal} {Physical Review B}\ }\textbf {\bibinfo
  {volume} {57}},\ \bibinfo {pages} {11039R} (\bibinfo {year}
  {1998})}\BibitemShut {NoStop}%
\bibitem [{\citenamefont {Fujiyama}\ \emph {et~al.}(2014)\citenamefont
  {Fujiyama}, \citenamefont {Ohsumi}, \citenamefont {Ohashi}, \citenamefont
  {Hirai}, \citenamefont {Kim}, \citenamefont {Arima}, \citenamefont {Takata},\
  and\ \citenamefont {Takagi}}]{Fujiyama:2014}%
  \BibitemOpen
  \bibfield  {author} {\bibinfo {author} {\bibfnamefont {S.}~\bibnamefont
  {Fujiyama}}, \bibinfo {author} {\bibfnamefont {H.}~\bibnamefont {Ohsumi}},
  \bibinfo {author} {\bibfnamefont {K.}~\bibnamefont {Ohashi}}, \bibinfo
  {author} {\bibfnamefont {D.}~\bibnamefont {Hirai}}, \bibinfo {author}
  {\bibfnamefont {B.~J.}\ \bibnamefont {Kim}}, \bibinfo {author} {\bibfnamefont
  {T.}~\bibnamefont {Arima}}, \bibinfo {author} {\bibfnamefont
  {M.}~\bibnamefont {Takata}}, \ and\ \bibinfo {author} {\bibfnamefont
  {H.}~\bibnamefont {Takagi}},\ }\href {\doibase
  10.1103/PhysRevLett.112.016405} {\bibfield  {journal} {\bibinfo  {journal}
  {Phys. Rev. Lett.}\ }\textbf {\bibinfo {volume} {112}},\ \bibinfo {pages}
  {016405} (\bibinfo {year} {2014})}\BibitemShut {NoStop}%
\bibitem [{\citenamefont {Kim}\ \emph {et~al.}(2012{\natexlab{a}})\citenamefont
  {Kim}, \citenamefont {Casa}, \citenamefont {Upton}, \citenamefont {Gog},
  \citenamefont {Kim}, \citenamefont {Mitchell}, \citenamefont {van
  Veenendaal}, \citenamefont {Daghofer}, \citenamefont {van~den Brink},
  \citenamefont {Khaliullin},\ and\ \citenamefont {Kim}}]{Kim:2012}%
  \BibitemOpen
  \bibfield  {author} {\bibinfo {author} {\bibfnamefont {J.}~\bibnamefont
  {Kim}}, \bibinfo {author} {\bibfnamefont {D.}~\bibnamefont {Casa}}, \bibinfo
  {author} {\bibfnamefont {M.~H.}\ \bibnamefont {Upton}}, \bibinfo {author}
  {\bibfnamefont {T.}~\bibnamefont {Gog}}, \bibinfo {author} {\bibfnamefont
  {Y.-J.}\ \bibnamefont {Kim}}, \bibinfo {author} {\bibfnamefont {J.~F.}\
  \bibnamefont {Mitchell}}, \bibinfo {author} {\bibfnamefont {M.}~\bibnamefont
  {van Veenendaal}}, \bibinfo {author} {\bibfnamefont {M.}~\bibnamefont
  {Daghofer}}, \bibinfo {author} {\bibfnamefont {J.}~\bibnamefont {van~den
  Brink}}, \bibinfo {author} {\bibfnamefont {G.}~\bibnamefont {Khaliullin}}, \
  and\ \bibinfo {author} {\bibfnamefont {B.~J.}\ \bibnamefont {Kim}},\ }\href
  {\doibase 10.1103/PhysRevLett.108.177003} {\bibfield  {journal} {\bibinfo
  {journal} {Phys. Rev. Lett.}\ }\textbf {\bibinfo {volume} {108}},\ \bibinfo
  {pages} {177003} (\bibinfo {year} {2012}{\natexlab{a}})}\BibitemShut
  {NoStop}%
\bibitem [{\citenamefont {Mitchell}(2015)}]{Mitchell:2015}%
  \BibitemOpen
  \bibfield  {author} {\bibinfo {author} {\bibfnamefont {J.~F.}\ \bibnamefont
  {Mitchell}},\ }\href {\doibase 10.1063/1.4921953} {\bibfield  {journal}
  {\bibinfo  {journal} {APL Mater.}\ }\textbf {\bibinfo {volume} {3}},\
  \bibinfo {pages} {062404} (\bibinfo {year} {2015})}\BibitemShut {NoStop}%
\bibitem [{\citenamefont {Di~Matteo}\ and\ \citenamefont
  {Norman}(2016)}]{Matteo:2016}%
  \BibitemOpen
  \bibfield  {author} {\bibinfo {author} {\bibfnamefont {S.}~\bibnamefont
  {Di~Matteo}}\ and\ \bibinfo {author} {\bibfnamefont {M.~R.}\ \bibnamefont
  {Norman}},\ }\href {\doibase 10.1103/PhysRevB.94.075148} {\bibfield
  {journal} {\bibinfo  {journal} {Phys. Rev. B}\ }\textbf {\bibinfo {volume}
  {94}},\ \bibinfo {pages} {075148} (\bibinfo {year} {2016})}\BibitemShut
  {NoStop}%
\bibitem [{\citenamefont {Sumita}\ \emph {et~al.}(2017)\citenamefont {Sumita},
  \citenamefont {Nomoto},\ and\ \citenamefont {Yanase}}]{Sumita:2017}%
  \BibitemOpen
  \bibfield  {author} {\bibinfo {author} {\bibfnamefont {S.}~\bibnamefont
  {Sumita}}, \bibinfo {author} {\bibfnamefont {T.}~\bibnamefont {Nomoto}}, \
  and\ \bibinfo {author} {\bibfnamefont {Y.}~\bibnamefont {Yanase}},\ }\href
  {\doibase 10.1103/PhysRevLett.119.027001} {\bibfield  {journal} {\bibinfo
  {journal} {Phys. Rev. Lett.}\ }\textbf {\bibinfo {volume} {119}},\ \bibinfo
  {pages} {027001} (\bibinfo {year} {2017})}\BibitemShut {NoStop}%
\bibitem [{\citenamefont {Porras}\ \emph {et~al.}(2019)\citenamefont {Porras},
  \citenamefont {Bertinshaw}, \citenamefont {Liu}, \citenamefont {Khaliullin},
  \citenamefont {Sung}, \citenamefont {Kim}, \citenamefont {Francoual},
  \citenamefont {Steffens}, \citenamefont {Deng}, \citenamefont {Sala},
  \citenamefont {Efimenko}, \citenamefont {Said}, \citenamefont {Casa},
  \citenamefont {Huang}, \citenamefont {Gog}, \citenamefont {Kim},
  \citenamefont {Keimer},\ and\ \citenamefont {Kim}}]{Porras:2019}%
  \BibitemOpen
  \bibfield  {author} {\bibinfo {author} {\bibfnamefont {J.}~\bibnamefont
  {Porras}}, \bibinfo {author} {\bibfnamefont {J.}~\bibnamefont {Bertinshaw}},
  \bibinfo {author} {\bibfnamefont {H.}~\bibnamefont {Liu}}, \bibinfo {author}
  {\bibfnamefont {G.}~\bibnamefont {Khaliullin}}, \bibinfo {author}
  {\bibfnamefont {N.~H.}\ \bibnamefont {Sung}}, \bibinfo {author}
  {\bibfnamefont {J.-W.}\ \bibnamefont {Kim}}, \bibinfo {author} {\bibfnamefont
  {S.}~\bibnamefont {Francoual}}, \bibinfo {author} {\bibfnamefont
  {P.}~\bibnamefont {Steffens}}, \bibinfo {author} {\bibfnamefont
  {G.}~\bibnamefont {Deng}}, \bibinfo {author} {\bibfnamefont {M.~M.}\
  \bibnamefont {Sala}}, \bibinfo {author} {\bibfnamefont {A.}~\bibnamefont
  {Efimenko}}, \bibinfo {author} {\bibfnamefont {A.}~\bibnamefont {Said}},
  \bibinfo {author} {\bibfnamefont {D.}~\bibnamefont {Casa}}, \bibinfo {author}
  {\bibfnamefont {X.}~\bibnamefont {Huang}}, \bibinfo {author} {\bibfnamefont
  {T.}~\bibnamefont {Gog}}, \bibinfo {author} {\bibfnamefont {J.}~\bibnamefont
  {Kim}}, \bibinfo {author} {\bibfnamefont {B.}~\bibnamefont {Keimer}}, \ and\
  \bibinfo {author} {\bibfnamefont {B.~J.}\ \bibnamefont {Kim}},\ }\href
  {\doibase 10.1103/PhysRevB.99.085125} {\bibfield  {journal} {\bibinfo
  {journal} {Phys. Rev. B}\ }\textbf {\bibinfo {volume} {99}},\ \bibinfo
  {pages} {085125} (\bibinfo {year} {2019})}\BibitemShut {NoStop}%
\bibitem [{\citenamefont {Fujiyama}\ \emph {et~al.}(2012)\citenamefont
  {Fujiyama}, \citenamefont {Ohsumi}, \citenamefont {Komesu}, \citenamefont
  {Matsuno}, \citenamefont {Kim}, \citenamefont {Takata}, \citenamefont
  {Arima},\ and\ \citenamefont {Takagi}}]{Fujiyama:2012}%
  \BibitemOpen
  \bibfield  {author} {\bibinfo {author} {\bibfnamefont {S.}~\bibnamefont
  {Fujiyama}}, \bibinfo {author} {\bibfnamefont {H.}~\bibnamefont {Ohsumi}},
  \bibinfo {author} {\bibfnamefont {T.}~\bibnamefont {Komesu}}, \bibinfo
  {author} {\bibfnamefont {J.}~\bibnamefont {Matsuno}}, \bibinfo {author}
  {\bibfnamefont {B.~J.}\ \bibnamefont {Kim}}, \bibinfo {author} {\bibfnamefont
  {M.}~\bibnamefont {Takata}}, \bibinfo {author} {\bibfnamefont
  {T.}~\bibnamefont {Arima}}, \ and\ \bibinfo {author} {\bibfnamefont
  {H.}~\bibnamefont {Takagi}},\ }\href {\doibase
  10.1103/PhysRevLett.108.247212} {\bibfield  {journal} {\bibinfo  {journal}
  {Phys. Rev. Lett.}\ }\textbf {\bibinfo {volume} {108}},\ \bibinfo {pages}
  {247212} (\bibinfo {year} {2012})}\BibitemShut {NoStop}%
\bibitem [{\citenamefont {Vale}\ \emph {et~al.}(2015)\citenamefont {Vale},
  \citenamefont {Boseggia}, \citenamefont {Walker}, \citenamefont {Springell},
  \citenamefont {Feng}, \citenamefont {Hunter}, \citenamefont {Perry},
  \citenamefont {Prabhakaran}, \citenamefont {Boothroyd}, \citenamefont
  {Collins}, \citenamefont {R\o{}nnow},\ and\ \citenamefont
  {McMorrow}}]{Vale:2015}%
  \BibitemOpen
  \bibfield  {author} {\bibinfo {author} {\bibfnamefont {J.~G.}\ \bibnamefont
  {Vale}}, \bibinfo {author} {\bibfnamefont {S.}~\bibnamefont {Boseggia}},
  \bibinfo {author} {\bibfnamefont {H.~C.}\ \bibnamefont {Walker}}, \bibinfo
  {author} {\bibfnamefont {R.}~\bibnamefont {Springell}}, \bibinfo {author}
  {\bibfnamefont {Z.}~\bibnamefont {Feng}}, \bibinfo {author} {\bibfnamefont
  {E.~C.}\ \bibnamefont {Hunter}}, \bibinfo {author} {\bibfnamefont {R.~S.}\
  \bibnamefont {Perry}}, \bibinfo {author} {\bibfnamefont {D.}~\bibnamefont
  {Prabhakaran}}, \bibinfo {author} {\bibfnamefont {A.~T.}\ \bibnamefont
  {Boothroyd}}, \bibinfo {author} {\bibfnamefont {S.~P.}\ \bibnamefont
  {Collins}}, \bibinfo {author} {\bibfnamefont {H.~M.}\ \bibnamefont
  {R\o{}nnow}}, \ and\ \bibinfo {author} {\bibfnamefont {D.~F.}\ \bibnamefont
  {McMorrow}},\ }\href {\doibase 10.1103/PhysRevB.92.020406} {\bibfield
  {journal} {\bibinfo  {journal} {Phys. Rev. B}\ }\textbf {\bibinfo {volume}
  {92}},\ \bibinfo {pages} {020406} (\bibinfo {year} {2015})}\BibitemShut
  {NoStop}%
\bibitem [{\citenamefont {Lovesey}\ \emph {et~al.}(2012)\citenamefont
  {Lovesey}, \citenamefont {Khalyavin}, \citenamefont {Manuel}, \citenamefont
  {Chapon}, \citenamefont {Cao},\ and\ \citenamefont {Qi}}]{Lovesey:2012}%
  \BibitemOpen
  \bibfield  {author} {\bibinfo {author} {\bibfnamefont {S.}~\bibnamefont
  {Lovesey}}, \bibinfo {author} {\bibfnamefont {D.}~\bibnamefont {Khalyavin}},
  \bibinfo {author} {\bibfnamefont {P.}~\bibnamefont {Manuel}}, \bibinfo
  {author} {\bibfnamefont {L.}~\bibnamefont {Chapon}}, \bibinfo {author}
  {\bibfnamefont {G.}~\bibnamefont {Cao}}, \ and\ \bibinfo {author}
  {\bibfnamefont {T.}~\bibnamefont {Qi}},\ }\href {\doibase
  10.1088/0953-8984/24/49/496003} {\bibfield  {journal} {\bibinfo  {journal}
  {J. Phys.: Condens. Matter}\ }\textbf {\bibinfo {volume} {24}},\ \bibinfo
  {pages} {496003} (\bibinfo {year} {2012})}\BibitemShut {NoStop}%
\bibitem [{\citenamefont {Kastner}\ \emph {et~al.}(1998)\citenamefont
  {Kastner}, \citenamefont {Birgeneau}, \citenamefont {Shirane},\ and\
  \citenamefont {Endoh}}]{Kastner:1998}%
  \BibitemOpen
  \bibfield  {author} {\bibinfo {author} {\bibfnamefont {M.~A.}\ \bibnamefont
  {Kastner}}, \bibinfo {author} {\bibfnamefont {R.~J.}\ \bibnamefont
  {Birgeneau}}, \bibinfo {author} {\bibfnamefont {G.}~\bibnamefont {Shirane}},
  \ and\ \bibinfo {author} {\bibfnamefont {Y.}~\bibnamefont {Endoh}},\ }\href
  {\doibase 10.1103/RevModPhys.70.897} {\bibfield  {journal} {\bibinfo
  {journal} {Rev. Mod. Phys.}\ }\textbf {\bibinfo {volume} {70}},\ \bibinfo
  {pages} {897} (\bibinfo {year} {1998})}\BibitemShut {NoStop}%
\bibitem [{\citenamefont {Haskel}\ \emph {et~al.}(2012)\citenamefont {Haskel},
  \citenamefont {Fabbris}, \citenamefont {Zhernenkov}, \citenamefont {Kong},
  \citenamefont {Jin}, \citenamefont {Cao},\ and\ \citenamefont {van
  Veenendaal}}]{Haskel:2012}%
  \BibitemOpen
  \bibfield  {author} {\bibinfo {author} {\bibfnamefont {D.}~\bibnamefont
  {Haskel}}, \bibinfo {author} {\bibfnamefont {G.}~\bibnamefont {Fabbris}},
  \bibinfo {author} {\bibfnamefont {M.}~\bibnamefont {Zhernenkov}}, \bibinfo
  {author} {\bibfnamefont {P.~P.}\ \bibnamefont {Kong}}, \bibinfo {author}
  {\bibfnamefont {C.~Q.}\ \bibnamefont {Jin}}, \bibinfo {author} {\bibfnamefont
  {G.}~\bibnamefont {Cao}}, \ and\ \bibinfo {author} {\bibfnamefont
  {M.}~\bibnamefont {van Veenendaal}},\ }\href {\doibase
  10.1103/PhysRevLett.109.027204} {\bibfield  {journal} {\bibinfo  {journal}
  {Phys. Rev. Lett.}\ }\textbf {\bibinfo {volume} {109}},\ \bibinfo {pages}
  {027204} (\bibinfo {year} {2012})}\BibitemShut {NoStop}%
\bibitem [{\citenamefont {Nichols}\ \emph {et~al.}(2013)\citenamefont
  {Nichols}, \citenamefont {Terzic}, \citenamefont {Bittle}, \citenamefont
  {Korneta}, \citenamefont {De~Long}, \citenamefont {Brill}, \citenamefont
  {Cao},\ and\ \citenamefont {Seo}}]{Nichols:2013}%
  \BibitemOpen
  \bibfield  {author} {\bibinfo {author} {\bibfnamefont {J.}~\bibnamefont
  {Nichols}}, \bibinfo {author} {\bibfnamefont {J.}~\bibnamefont {Terzic}},
  \bibinfo {author} {\bibfnamefont {E.~G.}\ \bibnamefont {Bittle}}, \bibinfo
  {author} {\bibfnamefont {O.~B.}\ \bibnamefont {Korneta}}, \bibinfo {author}
  {\bibfnamefont {L.~E.}\ \bibnamefont {De~Long}}, \bibinfo {author}
  {\bibfnamefont {J.~W.}\ \bibnamefont {Brill}}, \bibinfo {author}
  {\bibfnamefont {G.}~\bibnamefont {Cao}}, \ and\ \bibinfo {author}
  {\bibfnamefont {S.~S.~A.}\ \bibnamefont {Seo}},\ }\href {\doibase
  10.1063/1.4801877} {\bibfield  {journal} {\bibinfo  {journal} {Appl. Phys.
  Lett.}\ }\textbf {\bibinfo {volume} {102}},\ \bibinfo {pages} {141908}
  (\bibinfo {year} {2013})}\BibitemShut {NoStop}%
\bibitem [{\citenamefont {Serrao}\ \emph {et~al.}(2013)\citenamefont {Serrao},
  \citenamefont {Liu}, \citenamefont {Heron}, \citenamefont {G.Singh-Bhalla},
  \citenamefont {Yadav}, \citenamefont {Suresha}, \citenamefont {Suresha},
  \citenamefont {Paull}, \citenamefont {Yi}, \citenamefont {Chu}, \citenamefont
  {M.Trassin}, \citenamefont {Vishwanath}, \citenamefont {Arenholz},
  \citenamefont {Frontera}, \citenamefont {Zelezny}, \citenamefont {Jungwirth},
  \citenamefont {Marti},\ and\ \citenamefont {Ramesh}}]{Serrao:2013}%
  \BibitemOpen
  \bibfield  {author} {\bibinfo {author} {\bibfnamefont {C.~R.}\ \bibnamefont
  {Serrao}}, \bibinfo {author} {\bibfnamefont {J.}~\bibnamefont {Liu}},
  \bibinfo {author} {\bibfnamefont {J.}~\bibnamefont {Heron}}, \bibinfo
  {author} {\bibnamefont {G.Singh-Bhalla}}, \bibinfo {author} {\bibfnamefont
  {A.}~\bibnamefont {Yadav}}, \bibinfo {author} {\bibfnamefont
  {S.}~\bibnamefont {Suresha}}, \bibinfo {author} {\bibfnamefont
  {R.}~\bibnamefont {Suresha}}, \bibinfo {author} {\bibfnamefont
  {R.}~\bibnamefont {Paull}}, \bibinfo {author} {\bibfnamefont
  {D.}~\bibnamefont {Yi}}, \bibinfo {author} {\bibfnamefont {J.}~\bibnamefont
  {Chu}}, \bibinfo {author} {\bibnamefont {M.Trassin}}, \bibinfo {author}
  {\bibfnamefont {A.}~\bibnamefont {Vishwanath}}, \bibinfo {author}
  {\bibfnamefont {E.}~\bibnamefont {Arenholz}}, \bibinfo {author}
  {\bibfnamefont {C.}~\bibnamefont {Frontera}}, \bibinfo {author}
  {\bibfnamefont {J.}~\bibnamefont {Zelezny}}, \bibinfo {author} {\bibfnamefont
  {T.}~\bibnamefont {Jungwirth}}, \bibinfo {author} {\bibfnamefont
  {X.}~\bibnamefont {Marti}}, \ and\ \bibinfo {author} {\bibfnamefont
  {R.}~\bibnamefont {Ramesh}},\ }\href@noop {} {\bibfield  {journal} {\bibinfo
  {journal} {Phys. Rev. B}\ }\textbf {\bibinfo {volume} {87}},\ \bibinfo
  {pages} {085121} (\bibinfo {year} {2013})}\BibitemShut {NoStop}%
\bibitem [{\citenamefont {Lupascu}\ \emph {et~al.}(2014)\citenamefont
  {Lupascu}, \citenamefont {Clancy}, \citenamefont {Gretarsson}, \citenamefont
  {Nie}, \citenamefont {Nichols}, \citenamefont {Terzic}, \citenamefont {Cao},
  \citenamefont {Seo}, \citenamefont {Islam}, \citenamefont {Upton},
  \citenamefont {Kim}, \citenamefont {Casa}, \citenamefont {Gog}, \citenamefont
  {Said}, \citenamefont {Katukuri}, \citenamefont {Stoll}, \citenamefont
  {Hozoi}, \citenamefont {van~den Brink},\ and\ \citenamefont
  {Kim}}]{Lupascu:2014}%
  \BibitemOpen
  \bibfield  {author} {\bibinfo {author} {\bibfnamefont {A.}~\bibnamefont
  {Lupascu}}, \bibinfo {author} {\bibfnamefont {J.~P.}\ \bibnamefont {Clancy}},
  \bibinfo {author} {\bibfnamefont {H.}~\bibnamefont {Gretarsson}}, \bibinfo
  {author} {\bibfnamefont {Z.}~\bibnamefont {Nie}}, \bibinfo {author}
  {\bibfnamefont {J.}~\bibnamefont {Nichols}}, \bibinfo {author} {\bibfnamefont
  {J.}~\bibnamefont {Terzic}}, \bibinfo {author} {\bibfnamefont
  {G.}~\bibnamefont {Cao}}, \bibinfo {author} {\bibfnamefont {S.~S.~A.}\
  \bibnamefont {Seo}}, \bibinfo {author} {\bibfnamefont {Z.}~\bibnamefont
  {Islam}}, \bibinfo {author} {\bibfnamefont {M.~H.}\ \bibnamefont {Upton}},
  \bibinfo {author} {\bibfnamefont {J.}~\bibnamefont {Kim}}, \bibinfo {author}
  {\bibfnamefont {D.}~\bibnamefont {Casa}}, \bibinfo {author} {\bibfnamefont
  {T.}~\bibnamefont {Gog}}, \bibinfo {author} {\bibfnamefont {A.~H.}\
  \bibnamefont {Said}}, \bibinfo {author} {\bibfnamefont {V.~M.}\ \bibnamefont
  {Katukuri}}, \bibinfo {author} {\bibfnamefont {H.}~\bibnamefont {Stoll}},
  \bibinfo {author} {\bibfnamefont {L.}~\bibnamefont {Hozoi}}, \bibinfo
  {author} {\bibfnamefont {J.}~\bibnamefont {van~den Brink}}, \ and\ \bibinfo
  {author} {\bibfnamefont {Y.-J.}\ \bibnamefont {Kim}},\ }\href {\doibase
  10.1103/PhysRevLett.112.147201} {\bibfield  {journal} {\bibinfo  {journal}
  {Phys. Rev. Lett.}\ }\textbf {\bibinfo {volume} {112}},\ \bibinfo {pages}
  {147201} (\bibinfo {year} {2014})}\BibitemShut {NoStop}%
\bibitem [{\citenamefont {Miao}\ \emph {et~al.}(2014)\citenamefont {Miao},
  \citenamefont {Xu},\ and\ \citenamefont {Mao}}]{Miao:2014}%
  \BibitemOpen
  \bibfield  {author} {\bibinfo {author} {\bibfnamefont {L.}~\bibnamefont
  {Miao}}, \bibinfo {author} {\bibfnamefont {H.}~\bibnamefont {Xu}}, \ and\
  \bibinfo {author} {\bibfnamefont {Z.~Q.}\ \bibnamefont {Mao}},\ }\href
  {\doibase 10.1103/PhysRevB.89.035109} {\bibfield  {journal} {\bibinfo
  {journal} {Phys. Rev. B}\ }\textbf {\bibinfo {volume} {89}},\ \bibinfo
  {pages} {035109} (\bibinfo {year} {2014})}\BibitemShut {NoStop}%
\bibitem [{\citenamefont {Liu}\ \emph {et~al.}(2015)\citenamefont {Liu},
  \citenamefont {Khmelevskyi}, \citenamefont {Kim}, \citenamefont {Marsman},
  \citenamefont {Li}, \citenamefont {Chen}, \citenamefont {Sarma},
  \citenamefont {Kresse},\ and\ \citenamefont {Franchini}}]{Liu:2015}%
  \BibitemOpen
  \bibfield  {author} {\bibinfo {author} {\bibfnamefont {P.}~\bibnamefont
  {Liu}}, \bibinfo {author} {\bibfnamefont {S.}~\bibnamefont {Khmelevskyi}},
  \bibinfo {author} {\bibfnamefont {B.}~\bibnamefont {Kim}}, \bibinfo {author}
  {\bibfnamefont {M.}~\bibnamefont {Marsman}}, \bibinfo {author} {\bibfnamefont
  {D.}~\bibnamefont {Li}}, \bibinfo {author} {\bibfnamefont {X.-Q.}\
  \bibnamefont {Chen}}, \bibinfo {author} {\bibfnamefont {D.~D.}\ \bibnamefont
  {Sarma}}, \bibinfo {author} {\bibfnamefont {G.}~\bibnamefont {Kresse}}, \
  and\ \bibinfo {author} {\bibfnamefont {C.}~\bibnamefont {Franchini}},\ }\href
  {\doibase 10.1103/PhysRevB.92.054428} {\bibfield  {journal} {\bibinfo
  {journal} {Phys. Rev. B}\ }\textbf {\bibinfo {volume} {92}},\ \bibinfo
  {pages} {054428} (\bibinfo {year} {2015})}\BibitemShut {NoStop}%
\bibitem [{\citenamefont {Kim}\ \emph {et~al.}(2016{\natexlab{b}})\citenamefont
  {Kim}, \citenamefont {Kim}, \citenamefont {Kim},\ and\ \citenamefont
  {Min}}]{Kim:2016a}%
  \BibitemOpen
  \bibfield  {author} {\bibinfo {author} {\bibfnamefont {B.}~\bibnamefont
  {Kim}}, \bibinfo {author} {\bibfnamefont {B.~H.}\ \bibnamefont {Kim}},
  \bibinfo {author} {\bibfnamefont {K.}~\bibnamefont {Kim}}, \ and\ \bibinfo
  {author} {\bibfnamefont {B.~I.}\ \bibnamefont {Min}},\ }\href
  {https://doi.org/10.1038/srep27095} {\bibfield  {journal} {\bibinfo
  {journal} {Sci. Rep.}\ }\textbf {\bibinfo {volume} {6}},\ \bibinfo {pages}
  {27095} (\bibinfo {year} {2016}{\natexlab{b}})}\BibitemShut {NoStop}%
\bibitem [{\citenamefont {Kim}\ \emph {et~al.}(2017{\natexlab{a}})\citenamefont
  {Kim}, \citenamefont {Liu},\ and\ \citenamefont {Franchini}}]{Kim:2017}%
  \BibitemOpen
  \bibfield  {author} {\bibinfo {author} {\bibfnamefont {B.}~\bibnamefont
  {Kim}}, \bibinfo {author} {\bibfnamefont {P.}~\bibnamefont {Liu}}, \ and\
  \bibinfo {author} {\bibfnamefont {C.}~\bibnamefont {Franchini}},\ }\href
  {\doibase 10.1103/PhysRevB.95.115111} {\bibfield  {journal} {\bibinfo
  {journal} {Phys. Rev. B}\ }\textbf {\bibinfo {volume} {95}},\ \bibinfo
  {pages} {115111} (\bibinfo {year} {2017}{\natexlab{a}})}\BibitemShut
  {NoStop}%
\bibitem [{\citenamefont {Gutiérrez-Llorente}\ \emph
  {et~al.}(2018)\citenamefont {Gutiérrez-Llorente}, \citenamefont {Iglesias},
  \citenamefont {Rodríguez-González},\ and\ \citenamefont
  {Rivadulla}}]{Llorente:2018}%
  \BibitemOpen
  \bibfield  {author} {\bibinfo {author} {\bibfnamefont {A.}~\bibnamefont
  {Gutiérrez-Llorente}}, \bibinfo {author} {\bibfnamefont {L.}~\bibnamefont
  {Iglesias}}, \bibinfo {author} {\bibfnamefont {B.}~\bibnamefont
  {Rodríguez-González}}, \ and\ \bibinfo {author} {\bibfnamefont
  {F.}~\bibnamefont {Rivadulla}},\ }\href {\doibase 10.1063/1.5042836}
  {\bibfield  {journal} {\bibinfo  {journal} {APL Mater.}\ }\textbf {\bibinfo
  {volume} {6}},\ \bibinfo {pages} {091101} (\bibinfo {year}
  {2018})}\BibitemShut {NoStop}%
\bibitem [{\citenamefont {{Bhandari}}\ \emph {et~al.}(2018)\citenamefont
  {{Bhandari}}, \citenamefont {{Popovi{\'c}}},\ and\ \citenamefont
  {{Satpathy}}}]{Bhandari:2018}%
  \BibitemOpen
  \bibfield  {author} {\bibinfo {author} {\bibfnamefont {C.}~\bibnamefont
  {{Bhandari}}}, \bibinfo {author} {\bibfnamefont {Z.~S.}\ \bibnamefont
  {{Popovi{\'c}}}}, \ and\ \bibinfo {author} {\bibfnamefont {S.}~\bibnamefont
  {{Satpathy}}},\ }\href@noop {} {\bibfield  {journal} {\bibinfo  {journal}
  {ArXiv}\ } (\bibinfo {year} {2018})},\ \Eprint
  {http://arxiv.org/abs/1802.09719} {arXiv:1802.09719 [cond-mat.mtrl-sci]}
  \BibitemShut {NoStop}%
\bibitem [{\citenamefont {Seo}\ \emph {et~al.}(2019)\citenamefont {Seo},
  \citenamefont {Stavropoulos}, \citenamefont {Kim}, \citenamefont {F\"ursich},
  \citenamefont {Souri}, \citenamefont {Connell}, \citenamefont {Gretarsson},
  \citenamefont {Minola}, \citenamefont {Kee},\ and\ \citenamefont
  {Keimer}}]{Seo:2019}%
  \BibitemOpen
  \bibfield  {author} {\bibinfo {author} {\bibfnamefont {A.}~\bibnamefont
  {Seo}}, \bibinfo {author} {\bibfnamefont {P.~P.}\ \bibnamefont
  {Stavropoulos}}, \bibinfo {author} {\bibfnamefont {H.-H.}\ \bibnamefont
  {Kim}}, \bibinfo {author} {\bibfnamefont {K.}~\bibnamefont {F\"ursich}},
  \bibinfo {author} {\bibfnamefont {M.}~\bibnamefont {Souri}}, \bibinfo
  {author} {\bibfnamefont {J.~G.}\ \bibnamefont {Connell}}, \bibinfo {author}
  {\bibfnamefont {H.}~\bibnamefont {Gretarsson}}, \bibinfo {author}
  {\bibfnamefont {M.}~\bibnamefont {Minola}}, \bibinfo {author} {\bibfnamefont
  {H.~Y.}\ \bibnamefont {Kee}}, \ and\ \bibinfo {author} {\bibfnamefont
  {B.}~\bibnamefont {Keimer}},\ }\href {\doibase 10.1103/PhysRevB.100.165106}
  {\bibfield  {journal} {\bibinfo  {journal} {Phys. Rev. B}\ }\textbf {\bibinfo
  {volume} {100}},\ \bibinfo {pages} {165106} (\bibinfo {year}
  {2019})}\BibitemShut {NoStop}%
\bibitem [{\citenamefont {Carbone}(2004)}]{Carbone:2004}%
  \BibitemOpen
  \bibfield  {author} {\bibinfo {author} {\bibfnamefont {G.}~\bibnamefont
  {Carbone}},\ }\emph {\bibinfo {title} {Structural and magnetic studies of
  strained thin films of La$_{2/3}$Ca$_{1/3}$MnO$_{3}$}},\ \href@noop {} {Ph.D.
  thesis},\ \bibinfo  {school} {Department of Physics, University of Stuttgart}
  (\bibinfo {year} {2004})\BibitemShut {NoStop}%
\bibitem [{\citenamefont {Daumont}\ \emph {et~al.}(2009)\citenamefont
  {Daumont}, \citenamefont {Mannix}, \citenamefont {Venkatesan}, \citenamefont
  {Catalan}, \citenamefont {Ruby}, \citenamefont {Kool},\ and\ \citenamefont
  {Noheda}}]{Dumont:2009}%
  \BibitemOpen
  \bibfield  {author} {\bibinfo {author} {\bibfnamefont {C.}~\bibnamefont
  {Daumont}}, \bibinfo {author} {\bibfnamefont {D.}~\bibnamefont {Mannix}},
  \bibinfo {author} {\bibfnamefont {S.}~\bibnamefont {Venkatesan}}, \bibinfo
  {author} {\bibfnamefont {G.}~\bibnamefont {Catalan}}, \bibinfo {author}
  {\bibfnamefont {D.}~\bibnamefont {Ruby}}, \bibinfo {author} {\bibfnamefont
  {B.~J.}\ \bibnamefont {Kool}}, \ and\ \bibinfo {author} {\bibfnamefont {J.~T.
  M. D. H.~B.}\ \bibnamefont {Noheda}},\ }\href@noop {} {\bibfield  {journal}
  {\bibinfo  {journal} {J. Phys.: Condens. Matter}\ }\textbf {\bibinfo {volume}
  {81}},\ \bibinfo {pages} {182001} (\bibinfo {year} {2009})}\BibitemShut
  {NoStop}%
\bibitem [{\citenamefont {Opel}\ \emph {et~al.}(2013)\citenamefont {Opel},
  \citenamefont {Gepr\"{a}gs}, \citenamefont {Althammer}, \citenamefont
  {Brenninger},\ and\ \citenamefont {Gross}}]{Opel:2013}%
  \BibitemOpen
  \bibfield  {author} {\bibinfo {author} {\bibfnamefont {M.}~\bibnamefont
  {Opel}}, \bibinfo {author} {\bibfnamefont {S.}~\bibnamefont {Gepr\"{a}gs}},
  \bibinfo {author} {\bibfnamefont {M.}~\bibnamefont {Althammer}}, \bibinfo
  {author} {\bibfnamefont {T.}~\bibnamefont {Brenninger}}, \ and\ \bibinfo
  {author} {\bibfnamefont {R.}~\bibnamefont {Gross}},\ }\href {\doibase
  10.1088/0022-3727/47/3/034002} {\bibfield  {journal} {\bibinfo  {journal} {J.
  Phys. D: Appl. Phys.}\ }\textbf {\bibinfo {volume} {47}},\ \bibinfo {pages}
  {034002} (\bibinfo {year} {2013})}\BibitemShut {NoStop}%
\bibitem [{\citenamefont {Gross}\ \emph {et~al.}(2000)\citenamefont {Gross},
  \citenamefont {Klein}, \citenamefont {Wiedenhorst}, \citenamefont {Hoefener},
  \citenamefont {Schoop}, \citenamefont {Philipp}, \citenamefont {Schonecke},
  \citenamefont {Herbstritt}, \citenamefont {Alff}, \citenamefont {Lu},
  \citenamefont {Marx}, \citenamefont {Schymon}, \citenamefont {Thienhaus},\
  and\ \citenamefont {Mader}}]{Gross:2000}%
  \BibitemOpen
  \bibfield  {author} {\bibinfo {author} {\bibfnamefont {R.}~\bibnamefont
  {Gross}}, \bibinfo {author} {\bibfnamefont {J.}~\bibnamefont {Klein}},
  \bibinfo {author} {\bibfnamefont {B.}~\bibnamefont {Wiedenhorst}}, \bibinfo
  {author} {\bibfnamefont {C.}~\bibnamefont {Hoefener}}, \bibinfo {author}
  {\bibfnamefont {U.}~\bibnamefont {Schoop}}, \bibinfo {author} {\bibfnamefont
  {J.~B.}\ \bibnamefont {Philipp}}, \bibinfo {author} {\bibfnamefont
  {M.}~\bibnamefont {Schonecke}}, \bibinfo {author} {\bibfnamefont
  {F.}~\bibnamefont {Herbstritt}}, \bibinfo {author} {\bibfnamefont
  {L.}~\bibnamefont {Alff}}, \bibinfo {author} {\bibfnamefont {Y.}~\bibnamefont
  {Lu}}, \bibinfo {author} {\bibfnamefont {A.}~\bibnamefont {Marx}}, \bibinfo
  {author} {\bibfnamefont {S.}~\bibnamefont {Schymon}}, \bibinfo {author}
  {\bibfnamefont {S.}~\bibnamefont {Thienhaus}}, \ and\ \bibinfo {author}
  {\bibfnamefont {W.}~\bibnamefont {Mader}},\ }in\ \href {\doibase
  10.1117/12.397845} {\emph {\bibinfo {booktitle} {Superconducting and Related
  Oxides: Physics and Nanoengineering IV}}},\ Vol.\ \bibinfo {volume} {4058},\
  \bibinfo {editor} {edited by\ \bibinfo {editor} {\bibfnamefont
  {D.}~\bibnamefont {Pavuna}}\ and\ \bibinfo {editor} {\bibfnamefont
  {I.}~\bibnamefont {Bozovic}}},\ \bibinfo {organization} {International
  Society for Optics and Photonics}\ (\bibinfo  {publisher} {SPIE},\ \bibinfo
  {year} {2000})\ pp.\ \bibinfo {pages} {278 -- 294}\BibitemShut {NoStop}%
\bibitem [{\citenamefont {Bhatti}\ \emph {et~al.}(2014)\citenamefont {Bhatti},
  \citenamefont {Rawat}, \citenamefont {Banerjee},\ and\ \citenamefont
  {Pramanik}}]{Bhatti:2014}%
  \BibitemOpen
  \bibfield  {author} {\bibinfo {author} {\bibfnamefont {I.~N.}\ \bibnamefont
  {Bhatti}}, \bibinfo {author} {\bibfnamefont {R.}~\bibnamefont {Rawat}},
  \bibinfo {author} {\bibfnamefont {A.}~\bibnamefont {Banerjee}}, \ and\
  \bibinfo {author} {\bibfnamefont {A.~K.}\ \bibnamefont {Pramanik}},\ }\href
  {http://stacks.iop.org/0953-8984/27/i=1/a=016005} {\bibfield  {journal}
  {\bibinfo  {journal} {J. Phys.: Condens. Matter}\ }\textbf {\bibinfo {volume}
  {27}},\ \bibinfo {pages} {016005} (\bibinfo {year} {2014})}\BibitemShut
  {NoStop}%
\bibitem [{\citenamefont {Kim}\ \emph {et~al.}(2017{\natexlab{b}})\citenamefont
  {Kim}, \citenamefont {Liu},\ and\ \citenamefont {Franchini}}]{Kim:2017a}%
  \BibitemOpen
  \bibfield  {author} {\bibinfo {author} {\bibfnamefont {B.}~\bibnamefont
  {Kim}}, \bibinfo {author} {\bibfnamefont {P.}~\bibnamefont {Liu}}, \ and\
  \bibinfo {author} {\bibfnamefont {C.}~\bibnamefont {Franchini}},\ }\href
  {\doibase 10.1103/PhysRevB.95.024406} {\bibfield  {journal} {\bibinfo
  {journal} {Phys. Rev. B}\ }\textbf {\bibinfo {volume} {95}},\ \bibinfo
  {pages} {024406} (\bibinfo {year} {2017}{\natexlab{b}})}\BibitemShut
  {NoStop}%
\bibitem [{\citenamefont {Sung}\ \emph {et~al.}(2016)\citenamefont {Sung},
  \citenamefont {Gretarsson}, \citenamefont {Proepper}, \citenamefont {Porras},
  \citenamefont {Tacon}, \citenamefont {Boris}, \citenamefont {Keimer},\ and\
  \citenamefont {Kim}}]{Sung:2016}%
  \BibitemOpen
  \bibfield  {author} {\bibinfo {author} {\bibfnamefont {N.~H.}\ \bibnamefont
  {Sung}}, \bibinfo {author} {\bibfnamefont {H.}~\bibnamefont {Gretarsson}},
  \bibinfo {author} {\bibfnamefont {D.}~\bibnamefont {Proepper}}, \bibinfo
  {author} {\bibfnamefont {J.}~\bibnamefont {Porras}}, \bibinfo {author}
  {\bibfnamefont {M.~L.}\ \bibnamefont {Tacon}}, \bibinfo {author}
  {\bibfnamefont {A.~V.}\ \bibnamefont {Boris}}, \bibinfo {author}
  {\bibfnamefont {B.}~\bibnamefont {Keimer}}, \ and\ \bibinfo {author}
  {\bibfnamefont {B.~J.}\ \bibnamefont {Kim}},\ }\href {\doibase
  10.1080/14786435.2015.1134835} {\bibfield  {journal} {\bibinfo  {journal}
  {Philos. Mag.}\ }\textbf {\bibinfo {volume} {96}},\ \bibinfo {pages} {413}
  (\bibinfo {year} {2016})}\BibitemShut {NoStop}%
\bibitem [{\citenamefont {Afanasiev}\ \emph {et~al.}(2019)\citenamefont
  {Afanasiev}, \citenamefont {Gatilova}, \citenamefont {Groenendijk},
  \citenamefont {Ivanov}, \citenamefont {Gibert}, \citenamefont {Gariglio},
  \citenamefont {Mentink}, \citenamefont {Li}, \citenamefont {Dasari},
  \citenamefont {Eckstein}, \citenamefont {Rasing}, \citenamefont {Caviglia},\
  and\ \citenamefont {Kimel}}]{Afanasiev:2019}%
  \BibitemOpen
  \bibfield  {author} {\bibinfo {author} {\bibfnamefont {D.}~\bibnamefont
  {Afanasiev}}, \bibinfo {author} {\bibfnamefont {A.}~\bibnamefont {Gatilova}},
  \bibinfo {author} {\bibfnamefont {D.~J.}\ \bibnamefont {Groenendijk}},
  \bibinfo {author} {\bibfnamefont {B.~A.}\ \bibnamefont {Ivanov}}, \bibinfo
  {author} {\bibfnamefont {M.}~\bibnamefont {Gibert}}, \bibinfo {author}
  {\bibfnamefont {S.}~\bibnamefont {Gariglio}}, \bibinfo {author}
  {\bibfnamefont {J.}~\bibnamefont {Mentink}}, \bibinfo {author} {\bibfnamefont
  {J.}~\bibnamefont {Li}}, \bibinfo {author} {\bibfnamefont {N.}~\bibnamefont
  {Dasari}}, \bibinfo {author} {\bibfnamefont {M.}~\bibnamefont {Eckstein}},
  \bibinfo {author} {\bibfnamefont {T.}~\bibnamefont {Rasing}}, \bibinfo
  {author} {\bibfnamefont {A.~D.}\ \bibnamefont {Caviglia}}, \ and\ \bibinfo
  {author} {\bibfnamefont {A.~V.}\ \bibnamefont {Kimel}},\ }\href {\doibase
  10.1103/PhysRevX.9.021020} {\bibfield  {journal} {\bibinfo  {journal} {Phys.
  Rev. X}\ }\textbf {\bibinfo {volume} {9}},\ \bibinfo {pages} {021020}
  (\bibinfo {year} {2019})}\BibitemShut {NoStop}%
\bibitem [{\citenamefont {Okabe}\ \emph {et~al.}(2011)\citenamefont {Okabe},
  \citenamefont {Isobe}, \citenamefont {Takayama-Muromachi}, \citenamefont
  {Koda}, \citenamefont {Takeshita}, \citenamefont {Hiraishi}, \citenamefont
  {Miyazaki}, \citenamefont {Kadono}, \citenamefont {Miyake},\ and\
  \citenamefont {Akimitsu}}]{Okabe:2011}%
  \BibitemOpen
  \bibfield  {author} {\bibinfo {author} {\bibfnamefont {H.}~\bibnamefont
  {Okabe}}, \bibinfo {author} {\bibfnamefont {M.}~\bibnamefont {Isobe}},
  \bibinfo {author} {\bibfnamefont {E.}~\bibnamefont {Takayama-Muromachi}},
  \bibinfo {author} {\bibfnamefont {A.}~\bibnamefont {Koda}}, \bibinfo {author}
  {\bibfnamefont {S.}~\bibnamefont {Takeshita}}, \bibinfo {author}
  {\bibfnamefont {M.}~\bibnamefont {Hiraishi}}, \bibinfo {author}
  {\bibfnamefont {M.}~\bibnamefont {Miyazaki}}, \bibinfo {author}
  {\bibfnamefont {R.}~\bibnamefont {Kadono}}, \bibinfo {author} {\bibfnamefont
  {Y.}~\bibnamefont {Miyake}}, \ and\ \bibinfo {author} {\bibfnamefont
  {J.}~\bibnamefont {Akimitsu}},\ }\href {\doibase 10.1103/PhysRevB.83.155118}
  {\bibfield  {journal} {\bibinfo  {journal} {Phys. Rev. B}\ }\textbf {\bibinfo
  {volume} {83}},\ \bibinfo {pages} {155118} (\bibinfo {year}
  {2011})}\BibitemShut {NoStop}%
\bibitem [{\citenamefont {Guo}\ \emph {et~al.}(2020)\citenamefont {Guo},
  \citenamefont {Ji}, \citenamefont {Gu}, \citenamefont {Zhou}, \citenamefont
  {Nie},\ and\ \citenamefont {Pan}}]{Guo:2020}%
  \BibitemOpen
  \bibfield  {author} {\bibinfo {author} {\bibfnamefont {W.}~\bibnamefont
  {Guo}}, \bibinfo {author} {\bibfnamefont {D.~X.}\ \bibnamefont {Ji}},
  \bibinfo {author} {\bibfnamefont {Z.~B.}\ \bibnamefont {Gu}}, \bibinfo
  {author} {\bibfnamefont {J.}~\bibnamefont {Zhou}}, \bibinfo {author}
  {\bibfnamefont {Y.~F.}\ \bibnamefont {Nie}}, \ and\ \bibinfo {author}
  {\bibfnamefont {X.~Q.}\ \bibnamefont {Pan}},\ }\href {\doibase
  10.1103/PhysRevB.101.085101} {\bibfield  {journal} {\bibinfo  {journal}
  {Phys. Rev. B}\ }\textbf {\bibinfo {volume} {101}},\ \bibinfo {pages}
  {085101} (\bibinfo {year} {2020})}\BibitemShut {NoStop}%
\bibitem [{\citenamefont {Lu}\ \emph {et~al.}(2015)\citenamefont {Lu},
  \citenamefont {Dong}, \citenamefont {Quindeau}, \citenamefont {Preziosi},
  \citenamefont {Hu},\ and\ \citenamefont {Alexe}}]{Lu:2015}%
  \BibitemOpen
  \bibfield  {author} {\bibinfo {author} {\bibfnamefont {C.}~\bibnamefont
  {Lu}}, \bibinfo {author} {\bibfnamefont {S.}~\bibnamefont {Dong}}, \bibinfo
  {author} {\bibfnamefont {A.}~\bibnamefont {Quindeau}}, \bibinfo {author}
  {\bibfnamefont {D.}~\bibnamefont {Preziosi}}, \bibinfo {author}
  {\bibfnamefont {N.}~\bibnamefont {Hu}}, \ and\ \bibinfo {author}
  {\bibfnamefont {M.}~\bibnamefont {Alexe}},\ }\href {\doibase
  10.1103/PhysRevB.91.104401} {\bibfield  {journal} {\bibinfo  {journal} {Phys.
  Rev. B}\ }\textbf {\bibinfo {volume} {91}},\ \bibinfo {pages} {104401}
  (\bibinfo {year} {2015})}\BibitemShut {NoStop}%
\bibitem [{\citenamefont {Ge}\ \emph {et~al.}(2011)\citenamefont {Ge},
  \citenamefont {Qi}, \citenamefont {Korneta}, \citenamefont {De~Long},
  \citenamefont {Schlottmann}, \citenamefont {Crummett},\ and\ \citenamefont
  {Cao}}]{Ge:2011}%
  \BibitemOpen
  \bibfield  {author} {\bibinfo {author} {\bibfnamefont {M.}~\bibnamefont
  {Ge}}, \bibinfo {author} {\bibfnamefont {T.~F.}\ \bibnamefont {Qi}}, \bibinfo
  {author} {\bibfnamefont {O.~B.}\ \bibnamefont {Korneta}}, \bibinfo {author}
  {\bibfnamefont {D.~E.}\ \bibnamefont {De~Long}}, \bibinfo {author}
  {\bibfnamefont {P.}~\bibnamefont {Schlottmann}}, \bibinfo {author}
  {\bibfnamefont {W.~P.}\ \bibnamefont {Crummett}}, \ and\ \bibinfo {author}
  {\bibfnamefont {G.}~\bibnamefont {Cao}},\ }\href {\doibase
  10.1103/PhysRevB.84.100402} {\bibfield  {journal} {\bibinfo  {journal} {Phys.
  Rev. B}\ }\textbf {\bibinfo {volume} {84}},\ \bibinfo {pages} {100402}
  (\bibinfo {year} {2011})}\BibitemShut {NoStop}%
\bibitem [{\citenamefont {Shirane}\ and\ \citenamefont
  {Yamada}(1969)}]{Shirane:1969}%
  \BibitemOpen
  \bibfield  {author} {\bibinfo {author} {\bibfnamefont {G.}~\bibnamefont
  {Shirane}}\ and\ \bibinfo {author} {\bibfnamefont {Y.}~\bibnamefont
  {Yamada}},\ }\href {\doibase 10.1103/PhysRev.177.858} {\bibfield  {journal}
  {\bibinfo  {journal} {Phys. Rev.}\ }\textbf {\bibinfo {volume} {177}},\
  \bibinfo {pages} {858} (\bibinfo {year} {1969})}\BibitemShut {NoStop}%
\bibitem [{\citenamefont {Loetzsch}\ \emph {et~al.}(2010)\citenamefont
  {Loetzsch}, \citenamefont {Lübcke}, \citenamefont {Uschmann}, \citenamefont
  {Förster}, \citenamefont {Große}, \citenamefont {Thuerk}, \citenamefont
  {Koettig}, \citenamefont {Schmidl},\ and\ \citenamefont
  {Seidel}}]{Loetzsch:2010}%
  \BibitemOpen
  \bibfield  {author} {\bibinfo {author} {\bibfnamefont {R.}~\bibnamefont
  {Loetzsch}}, \bibinfo {author} {\bibfnamefont {A.}~\bibnamefont {Lübcke}},
  \bibinfo {author} {\bibfnamefont {I.}~\bibnamefont {Uschmann}}, \bibinfo
  {author} {\bibfnamefont {E.}~\bibnamefont {Förster}}, \bibinfo {author}
  {\bibfnamefont {V.}~\bibnamefont {Große}}, \bibinfo {author} {\bibfnamefont
  {M.}~\bibnamefont {Thuerk}}, \bibinfo {author} {\bibfnamefont
  {T.}~\bibnamefont {Koettig}}, \bibinfo {author} {\bibfnamefont
  {F.}~\bibnamefont {Schmidl}}, \ and\ \bibinfo {author} {\bibfnamefont
  {P.}~\bibnamefont {Seidel}},\ }\href {\doibase 10.1063/1.3324695} {\bibfield
  {journal} {\bibinfo  {journal} {Applied Physics Letters}\ }\textbf {\bibinfo
  {volume} {96}},\ \bibinfo {pages} {071901} (\bibinfo {year}
  {2010})}\BibitemShut {NoStop}%
\bibitem [{\citenamefont {Chahine}\ \emph {et~al.}(2019)\citenamefont
  {Chahine}, \citenamefont {Blanc}, \citenamefont {Arnaud}, \citenamefont
  {De~Geuser}, \citenamefont {Guinebreti\`{e}re},\ and\ \citenamefont
  {Boudet}}]{CRG-D2AM}%
  \BibitemOpen
  \bibfield  {author} {\bibinfo {author} {\bibfnamefont {G.~A.}\ \bibnamefont
  {Chahine}}, \bibinfo {author} {\bibfnamefont {N.}~\bibnamefont {Blanc}},
  \bibinfo {author} {\bibfnamefont {S.}~\bibnamefont {Arnaud}}, \bibinfo
  {author} {\bibfnamefont {F.}~\bibnamefont {De~Geuser}}, \bibinfo {author}
  {\bibfnamefont {R.}~\bibnamefont {Guinebreti\`{e}re}}, \ and\ \bibinfo
  {author} {\bibfnamefont {N.}~\bibnamefont {Boudet}},\ }\href {\doibase
  10.3390/met9030352} {\bibfield  {journal} {\bibinfo  {journal} {Metals}\ }\textbf {\bibinfo {volume} {9}},\
  \bibinfo {pages} {352} (\bibinfo {year} {2019})}\BibitemShut {NoStop}%
\bibitem [{\citenamefont {Katukuri}\ \emph {et~al.}(2012)\citenamefont
  {Katukuri}, \citenamefont {Stoll}, \citenamefont {van~den Brink},\ and\
  \citenamefont {Hozoi}}]{Katukuri:2012}%
  \BibitemOpen
  \bibfield  {author} {\bibinfo {author} {\bibfnamefont {V.~M.}\ \bibnamefont
  {Katukuri}}, \bibinfo {author} {\bibfnamefont {H.}~\bibnamefont {Stoll}},
  \bibinfo {author} {\bibfnamefont {J.}~\bibnamefont {van~den Brink}}, \ and\
  \bibinfo {author} {\bibfnamefont {L.}~\bibnamefont {Hozoi}},\ }\href
  {\doibase 10.1103/PhysRevB.85.220402} {\bibfield  {journal} {\bibinfo
  {journal} {Phys. Rev. B}\ }\textbf {\bibinfo {volume} {85}},\ \bibinfo
  {pages} {220402} (\bibinfo {year} {2012})}\BibitemShut {NoStop}%
\bibitem [{\citenamefont {Moretti~Sala}\ \emph
  {et~al.}(2014{\natexlab{a}})\citenamefont {Moretti~Sala}, \citenamefont
  {Rossi}, \citenamefont {Al-Zein}, \citenamefont {Boseggia}, \citenamefont
  {Hunter}, \citenamefont {Perry}, \citenamefont {Prabhakaran}, \citenamefont
  {Boothroyd}, \citenamefont {Brookes}, \citenamefont {McMorrow}, \citenamefont
  {Monaco},\ and\ \citenamefont {Krisch}}]{Sala:2014}%
  \BibitemOpen
  \bibfield  {author} {\bibinfo {author} {\bibfnamefont {M.}~\bibnamefont
  {Moretti~Sala}}, \bibinfo {author} {\bibfnamefont {M.}~\bibnamefont {Rossi}},
  \bibinfo {author} {\bibfnamefont {A.}~\bibnamefont {Al-Zein}}, \bibinfo
  {author} {\bibfnamefont {S.}~\bibnamefont {Boseggia}}, \bibinfo {author}
  {\bibfnamefont {E.~C.}\ \bibnamefont {Hunter}}, \bibinfo {author}
  {\bibfnamefont {R.~S.}\ \bibnamefont {Perry}}, \bibinfo {author}
  {\bibfnamefont {D.}~\bibnamefont {Prabhakaran}}, \bibinfo {author}
  {\bibfnamefont {A.~T.}\ \bibnamefont {Boothroyd}}, \bibinfo {author}
  {\bibfnamefont {N.~B.}\ \bibnamefont {Brookes}}, \bibinfo {author}
  {\bibfnamefont {D.~F.}\ \bibnamefont {McMorrow}}, \bibinfo {author}
  {\bibfnamefont {G.}~\bibnamefont {Monaco}}, \ and\ \bibinfo {author}
  {\bibfnamefont {M.}~\bibnamefont {Krisch}},\ }\href {\doibase
  10.1103/PhysRevB.90.085126} {\bibfield  {journal} {\bibinfo  {journal} {Phys.
  Rev. B}\ }\textbf {\bibinfo {volume} {90}},\ \bibinfo {pages} {085126}
  (\bibinfo {year} {2014}{\natexlab{a}})}\BibitemShut {NoStop}%
\bibitem [{\citenamefont {Kim}\ \emph {et~al.}(2012{\natexlab{b}})\citenamefont
  {Kim}, \citenamefont {Choi}, \citenamefont {Kim}, \citenamefont {Mitchell},
  \citenamefont {Jackeli}, \citenamefont {Daghofer}, \citenamefont {van~den
  Brink}, \citenamefont {Khaliullin},\ and\ \citenamefont {Kim}}]{Kim:2012b}%
  \BibitemOpen
  \bibfield  {author} {\bibinfo {author} {\bibfnamefont {J.~W.}\ \bibnamefont
  {Kim}}, \bibinfo {author} {\bibfnamefont {Y.}~\bibnamefont {Choi}}, \bibinfo
  {author} {\bibfnamefont {J.}~\bibnamefont {Kim}}, \bibinfo {author}
  {\bibfnamefont {J.~F.}\ \bibnamefont {Mitchell}}, \bibinfo {author}
  {\bibfnamefont {G.}~\bibnamefont {Jackeli}}, \bibinfo {author} {\bibfnamefont
  {M.}~\bibnamefont {Daghofer}}, \bibinfo {author} {\bibfnamefont
  {J.}~\bibnamefont {van~den Brink}}, \bibinfo {author} {\bibfnamefont
  {G.}~\bibnamefont {Khaliullin}}, \ and\ \bibinfo {author} {\bibfnamefont
  {B.~J.}\ \bibnamefont {Kim}},\ }\href {\doibase
  10.1103/PhysRevLett.109.037204} {\bibfield  {journal} {\bibinfo  {journal}
  {Phys. Rev. Lett.}\ }\textbf {\bibinfo {volume} {109}},\ \bibinfo {pages}
  {037204} (\bibinfo {year} {2012}{\natexlab{b}})}\BibitemShut {NoStop}%
\bibitem [{\citenamefont {Calder}\ \emph {et~al.}(2012)\citenamefont {Calder},
  \citenamefont {Cao}, \citenamefont {Lumsden}, \citenamefont {Kim},
  \citenamefont {Gai}, \citenamefont {Sales}, \citenamefont {Mandrus},\ and\
  \citenamefont {Christianson}}]{Calder:2012}%
  \BibitemOpen
  \bibfield  {author} {\bibinfo {author} {\bibfnamefont {S.}~\bibnamefont
  {Calder}}, \bibinfo {author} {\bibfnamefont {G.-X.}\ \bibnamefont {Cao}},
  \bibinfo {author} {\bibfnamefont {M.~D.}\ \bibnamefont {Lumsden}}, \bibinfo
  {author} {\bibfnamefont {J.~W.}\ \bibnamefont {Kim}}, \bibinfo {author}
  {\bibfnamefont {Z.}~\bibnamefont {Gai}}, \bibinfo {author} {\bibfnamefont
  {B.~C.}\ \bibnamefont {Sales}}, \bibinfo {author} {\bibfnamefont
  {D.}~\bibnamefont {Mandrus}}, \ and\ \bibinfo {author} {\bibfnamefont
  {A.~D.}\ \bibnamefont {Christianson}},\ }\href {\doibase
  10.1103/PhysRevB.86.220403} {\bibfield  {journal} {\bibinfo  {journal} {Phys.
  Rev. B}\ }\textbf {\bibinfo {volume} {86}},\ \bibinfo {pages} {220403}
  (\bibinfo {year} {2012})}\BibitemShut {NoStop}%
\bibitem [{\citenamefont {Ohgushi}\ \emph {et~al.}(2013)\citenamefont
  {Ohgushi}, \citenamefont {Yamaura}, \citenamefont {Ohsumi}, \citenamefont
  {Sugimoto}, \citenamefont {Takeshita}, \citenamefont {Tokuda}, \citenamefont
  {Takagi}, \citenamefont {Takata},\ and\ \citenamefont
  {Arima}}]{Ohgushi:2013}%
  \BibitemOpen
  \bibfield  {author} {\bibinfo {author} {\bibfnamefont {K.}~\bibnamefont
  {Ohgushi}}, \bibinfo {author} {\bibfnamefont {J.-i.}\ \bibnamefont
  {Yamaura}}, \bibinfo {author} {\bibfnamefont {H.}~\bibnamefont {Ohsumi}},
  \bibinfo {author} {\bibfnamefont {K.}~\bibnamefont {Sugimoto}}, \bibinfo
  {author} {\bibfnamefont {S.}~\bibnamefont {Takeshita}}, \bibinfo {author}
  {\bibfnamefont {A.}~\bibnamefont {Tokuda}}, \bibinfo {author} {\bibfnamefont
  {H.}~\bibnamefont {Takagi}}, \bibinfo {author} {\bibfnamefont
  {M.}~\bibnamefont {Takata}}, \ and\ \bibinfo {author} {\bibfnamefont {T.-h.}\
  \bibnamefont {Arima}},\ }\href {\doibase 10.1103/PhysRevLett.110.217212}
  {\bibfield  {journal} {\bibinfo  {journal} {Phys. Rev. Lett.}\ }\textbf
  {\bibinfo {volume} {110}},\ \bibinfo {pages} {217212} (\bibinfo {year}
  {2013})}\BibitemShut {NoStop}%
\bibitem [{\citenamefont {Boseggia}\ \emph
  {et~al.}(2013{\natexlab{b}})\citenamefont {Boseggia}, \citenamefont
  {Springell}, \citenamefont {Walker}, \citenamefont {R\o{}nnow}, \citenamefont
  {R\"uegg}, \citenamefont {Okabe}, \citenamefont {Isobe}, \citenamefont
  {Perry}, \citenamefont {Collins},\ and\ \citenamefont
  {McMorrow}}]{Boseggia:2013a}%
  \BibitemOpen
  \bibfield  {author} {\bibinfo {author} {\bibfnamefont {S.}~\bibnamefont
  {Boseggia}}, \bibinfo {author} {\bibfnamefont {R.}~\bibnamefont {Springell}},
  \bibinfo {author} {\bibfnamefont {H.~C.}\ \bibnamefont {Walker}}, \bibinfo
  {author} {\bibfnamefont {H.~M.}\ \bibnamefont {R\o{}nnow}}, \bibinfo {author}
  {\bibfnamefont {C.}~\bibnamefont {R\"uegg}}, \bibinfo {author} {\bibfnamefont
  {H.}~\bibnamefont {Okabe}}, \bibinfo {author} {\bibfnamefont
  {M.}~\bibnamefont {Isobe}}, \bibinfo {author} {\bibfnamefont {R.~S.}\
  \bibnamefont {Perry}}, \bibinfo {author} {\bibfnamefont {S.~P.}\ \bibnamefont
  {Collins}}, \ and\ \bibinfo {author} {\bibfnamefont {D.~F.}\ \bibnamefont
  {McMorrow}},\ }\href {\doibase 10.1103/PhysRevLett.110.117207} {\bibfield
  {journal} {\bibinfo  {journal} {Phys. Rev. Lett.}\ }\textbf {\bibinfo
  {volume} {110}},\ \bibinfo {pages} {117207} (\bibinfo {year}
  {2013}{\natexlab{b}})}\BibitemShut {NoStop}%
\bibitem [{\citenamefont {Matteo}(2012)}]{Matteo:2012}%
  \BibitemOpen
  \bibfield  {author} {\bibinfo {author} {\bibfnamefont {S.~D.}\ \bibnamefont
  {Matteo}},\ }\href {http://stacks.iop.org/0022-3727/45/i=16/a=163001}
  {\bibfield  {journal} {\bibinfo  {journal} {J. Phys. D: Appl. Phys.}\
  }\textbf {\bibinfo {volume} {45}},\ \bibinfo {pages} {163001} (\bibinfo
  {year} {2012})}\BibitemShut {NoStop}%
\bibitem [{\citenamefont {Chapon}\ and\ \citenamefont
  {Lovesey}(2011)}]{Chapon:2011}%
  \BibitemOpen
  \bibfield  {author} {\bibinfo {author} {\bibfnamefont {L.~C.}\ \bibnamefont
  {Chapon}}\ and\ \bibinfo {author} {\bibfnamefont {S.~W.}\ \bibnamefont
  {Lovesey}},\ }\href {http://stacks.iop.org/0953-8984/23/i=25/a=252201}
  {\bibfield  {journal} {\bibinfo  {journal} {J. Phys.: Condens. Matter}\
  }\textbf {\bibinfo {volume} {23}},\ \bibinfo {pages} {252201} (\bibinfo
  {year} {2011})}\BibitemShut {NoStop}%
\bibitem [{\citenamefont {Moretti~Sala}\ \emph
  {et~al.}(2014{\natexlab{b}})\citenamefont {Moretti~Sala}, \citenamefont
  {Boseggia}, \citenamefont {McMorrow},\ and\ \citenamefont
  {Monaco}}]{Sala:2014a}%
  \BibitemOpen
  \bibfield  {author} {\bibinfo {author} {\bibfnamefont {M.}~\bibnamefont
  {Moretti~Sala}}, \bibinfo {author} {\bibfnamefont {S.}~\bibnamefont
  {Boseggia}}, \bibinfo {author} {\bibfnamefont {D.~F.}\ \bibnamefont
  {McMorrow}}, \ and\ \bibinfo {author} {\bibfnamefont {G.}~\bibnamefont
  {Monaco}},\ }\href {\doibase 10.1103/PhysRevLett.112.026403} {\bibfield
  {journal} {\bibinfo  {journal} {Phys. Rev. Lett.}\ }\textbf {\bibinfo
  {volume} {112}},\ \bibinfo {pages} {026403} (\bibinfo {year}
  {2014}{\natexlab{b}})}\BibitemShut {NoStop}%
\bibitem [{\citenamefont {Stanley}(1971)}]{Stanley:1971}%
  \BibitemOpen
  \bibfield  {author} {\bibinfo {author} {\bibfnamefont {H.~E.}\ \bibnamefont
  {Stanley}},\ }\href@noop {} {\emph {\bibinfo {title} {"Introduction to Phase
  Transitions and Critical Phenomena"}}}\ (\bibinfo  {publisher} {Oxford
  University Press},\ \bibinfo {year} {1971})\BibitemShut {NoStop}%
\bibitem [{\citenamefont {Bramwell}\ and\ \citenamefont
  {Holdsworth}(1993)}]{Bramwell:1993}%
  \BibitemOpen
  \bibfield  {author} {\bibinfo {author} {\bibfnamefont {S.~T.}\ \bibnamefont
  {Bramwell}}\ and\ \bibinfo {author} {\bibfnamefont {P.~C.~W.}\ \bibnamefont
  {Holdsworth}},\ }\href {http://stacks.iop.org/0953-8984/5/i=4/a=004}
  {\bibfield  {journal} {\bibinfo  {journal} {J. Phys.: Condens. Matter}\
  }\textbf {\bibinfo {volume} {5}},\ \bibinfo {pages} {L53} (\bibinfo {year}
  {1993})}\BibitemShut {NoStop}%
\bibitem [{\citenamefont {Taroni}\ \emph {et~al.}(2008)\citenamefont {Taroni},
  \citenamefont {Bramwell},\ and\ \citenamefont {Holdsworth}}]{Taroni:2008}%
  \BibitemOpen
  \bibfield  {author} {\bibinfo {author} {\bibfnamefont {A.}~\bibnamefont
  {Taroni}}, \bibinfo {author} {\bibfnamefont {S.~T.}\ \bibnamefont
  {Bramwell}}, \ and\ \bibinfo {author} {\bibfnamefont {P.~C.~W.}\ \bibnamefont
  {Holdsworth}},\ }\href {http://stacks.iop.org/0953-8984/20/i=27/a=275233}
  {\bibfield  {journal} {\bibinfo  {journal} {J. Phys.: Condens. Matter}\
  }\textbf {\bibinfo {volume} {20}},\ \bibinfo {pages} {275233} (\bibinfo
  {year} {2008})}\BibitemShut {NoStop}%
\bibitem [{\citenamefont {Hill}\ and\ \citenamefont
  {McMorrow}(1996)}]{Hill:1996}%
  \BibitemOpen
  \bibfield  {author} {\bibinfo {author} {\bibfnamefont {J.~P.}\ \bibnamefont
  {Hill}}\ and\ \bibinfo {author} {\bibfnamefont {D.~F.}\ \bibnamefont
  {McMorrow}},\ }\href {\doibase 10.1107/S0108767395012670} {\bibfield
  {journal} {\bibinfo  {journal} {Acta Crystallogr.}\ }\textbf {\bibinfo
  {volume} {A52}},\ \bibinfo {pages} {236} (\bibinfo {year}
  {1996})}\BibitemShut {NoStop}%
\end{thebibliography}

%

\end{document}